\documentclass[prb,reprint,a4paper,superscriptaddress,citeautoscript,floatfix]{revtex4-2}

\pdfoutput=1
\usepackage[charter]{mathdesign}
\usepackage{amsmath}

\usepackage[pdftex]{graphicx}
\usepackage{microtype}
\usepackage{bm}\let\vec\bm

\usepackage[svgnames]{xcolor}
\usepackage[
	colorlinks=True,linkcolor=DarkRed,citecolor=ForestGreen,urlcolor=MediumBlue,
	pdfstartview=FitH,bookmarks=False,pdfpagemode=UseNone
]{hyperref}

\usepackage{feynmp-auto}
\newcommand{\fmfstyle}{%
	\fmfpen{0.5pt}\fmfset{arrow_ang}{15}\fmfset{arrow_len}{2mm}%
	\fmfset{wiggly_len}{2mm}\fmfset{dot_size}{0.5thick}%
	\fmfset{curly_len}{2mm}\fmfset{zigzag_width}{1mm}%
}

\def\CG{C_{\mathrm{G}}}
\def\CQ{C_{\mathrm{Q}}}
\def\Ck{C_{\mathrm{kin},1}}
\def\dQ{d_{\mathrm{Q}}}

\begin{document}

\title{Theory of cross quantum capacitance}

\author{Christophe Berthod}
\affiliation{Department of Quantum Matter Physics, University of Geneva, 24 quai Ernest-Ansermet, 1211 Geneva, Switzerland}
\author{Haijing Zhang}
\affiliation{Department of Quantum Matter Physics, University of Geneva, 24 quai Ernest-Ansermet, 1211 Geneva, Switzerland}
\affiliation{Group of Applied Physics, University of Geneva, 24 quai Ernest-Ansermet, 1211 Geneva, Switzerland}
\affiliation{Max-Planck-Institute for Chemical Physics of Solids, N{\"{o}}thnitzer Strasse 40, Dresden, Germany}
\author{Alberto F. Morpurgo}
\affiliation{Department of Quantum Matter Physics, University of Geneva, 24 quai Ernest-Ansermet, 1211 Geneva, Switzerland}
\affiliation{Group of Applied Physics, University of Geneva, 24 quai Ernest-Ansermet, 1211 Geneva, Switzerland}
\author{Thierry Giamarchi}
\affiliation{Department of Quantum Matter Physics, University of Geneva, 24 quai Ernest-Ansermet, 1211 Geneva, Switzerland}

\date{August 26, 2021}

\begin{abstract}

Impressive progress in the control of atomically thin crystals is now enabling the realization of gated structures in which two electrodes are separated by atomic scale distances. The electrical capacitance of these structures is determined by phenomena that are not relevant in capacitors with larger electrode separation. With the aim to analyze these phenomena, we use linear response theory to develop a systematic description of capacitance for two coupled electron liquids, accounting for the wave nature of electrons, as well as for the effect of both intra and interlayer Coulomb interactions. Our theory leads to a general expression for the electrical capacitance in terms of both intra and interlayer electronic polarizabilities. The intralayer polarizability is directly related to the conventional expression for the quantum capacitance, whereas the interlayer polarizability term accounts for interaction-induced correlations between charges hosted by opposite capacitor plates. We refer to this latter term---which has not been considered earlier---as to the cross quantum capacitance. We discuss the implications of the general expression for the capacitance, show that it leads to established results when the effect of interlayer correlations is negligible, and that the intra and interlayer polarizabilities play a comparable role for capacitors with very small electrode separation (i.e., cross quantum capacitance effects can be large and cannot be neglected). Using two different approaches, we calculate the capacitance in specific cases, and find that the interlayer polarizability can be either positive or negative, so that---depending on the regime considered---the cross quantum capacitance can either increase or decrease the total capacitance. We conclude by showing that the cross quantum capacitance term can lead to a nonmonotonic evolution of the total capacitance with increasing separation between the capacitor plates, which would represent an unambiguous manifestation of the cross quantum capacitance if observed experimentally. 

\end{abstract}

\maketitle

\section{Introduction}

The electrical capacitance describes the accumulation of charge on conductors in response to an applied potential difference and plays a key role in determining the electronic properties of many nanostructures \cite{Jacak-1998, Yu-2013, Zhu-2014, Chen-2015}. It is long known that the capacitance of nanostructures differs from the so-called geometrical capacitance calculated using the equations of classical electromagnetism, which assume that all conducting electrodes are ideal metals and have an infinite density of states \cite{Luryi-1988}. Indeed, the geometrical capacitance neglects all material specific properties, that in practice determine how much charge can be accumulated. To account for material dependent properties, a variety of phenomena and mechanisms have been considered by theory when analyzing the capacitance measured in experiments \cite{Eisenstein-1992, Xia-2009, Li-2011, Yu-2013, Zhu-2014, Chen-2015}.

In the majority of cases, the capacitance is calculated theoretically starting from the energy of the system as a function of accumulated charge density $n$ and taking the second derivative, using the relation \cite{Kopp-2009}
	\begin{equation}\label{eq:CE}
		\frac{1}{C}=\frac{1}{e^2}\frac{d^2E(n)}{dn^2},
	\end{equation}
exactly valid for the geometrical capacitance in classical electromagnetism. However, as the energy of virtually all electronic systems of interest cannot be calculated exactly, approximations are needed and the results depend on which contributions to $E(n)$ are considered. Most past theoretical work has followed one of two distinct approaches.

In the first approach, material properties are analyzed first to determine the specific behavior of the two conductors used as capacitor plates, which is then taken into account when calculating the electrostatic energy. For instance, in capacitors formed by conductors separated by a distance smaller than the interparticle separation in the electrodes, it may happen that positive and negative charges on opposite plates spontaneously bind to form neutral excitons \cite{Skinner-2010}. Charging the capacitor requires overcoming the repulsion between excitons that interact through dipolar forces, i.e., much more weakly than unpaired charges. As a result, the energy needed to accumulate a given exciton density is smaller than that needed to accumulate the same density of unpaired positive and negative charges, resulting in a capacitance larger than the geometrical one. Similarly, materials forming the electrodes may exhibit a spontaneous modulation of the charge density---such as in a Wigner crystal---causing the electrostatic energy to differ from that of a classical capacitor, in which the electron density is uniform \cite{Chitra-2005-1}. The energy considered in the calculation of the capacitance in this first approach, therefore, is due to the Coulomb interaction of a classical distribution of charges on opposite plates, and the result deviates from the geometrical capacitance because this distribution can be nonuniform. In most cases, the capacitance is found to be larger than the geometrical capacitance \cite{Skinner-2010, Skinner-2013-1, Skinner-2013-2, Chitra-2005-1}.

The second approach that is followed frequently relies on the concept of quantum capacitance \cite{Giuliani-Vignale}. In this case, a system of interest is used as plate of a capacitor and the other plate is considered to play no role in the calculation of the total energy (as if it was an ideal metal in the sense of classical electrostatics). The energy $E(n)$ used to extract the capacitance is then the sum of the classical electrostatic energy (i.e., the energy stored in the electric field inside the capacitor) and the change in the energy of the system of interest, upon varying $n$ \cite{Giuliani-Vignale}. In the simplest case, the dominant energy term is the increase in kinetic energy of the added electrons. It is straightforward to show that this term gives a contribution---the quantum capacitance---proportional to the density of states, which adds in series to the geometrical capacitance, so that the total capacitance is reduced \cite{Luryi-1988}. The situation can be richer if interactions play an important role, since then the quantum capacitance is proportional to the electronic compressibility of the system of interest \cite{Steffen-2017}. Some past experiments, for instance, have been interpreted in terms of negative compressibility due to exchange interaction, resulting in a capacitance larger than the geometrical one \cite{Eisenstein-1992, Li-2011}.

The differences between these two approaches are striking. In the first one, deviations from the geometrical capacitance are determined by the energy associated to interactions between charges on the two capacitor plates. They are calculated considering the electrostatic interaction of a classical charge distribution, treating charge carriers as localized particles, neglecting important aspects of the physics associated to the wave nature of electrons. In the second approach, instead, deviations from the geometrical capacitance due to interactions between charges on the two plates are entirely disregarded. In this case, it is the wave nature of the electrons that plays a central role, which is why interesting physics in the system of interest is associated to the finite density of states and to the effect of exchange interactions. It seems obvious that in capacitors whose plates are separated by very short distances, both the effect of interplate interactions and the wave nature of electrons should be considered simultaneously. However, the difference in the theoretical approaches followed in the past---as well as in the methods used to implement them---make it difficult to identify a suitable formalism for a proper theoretical description.

Here we propose a first systematic theoretical treatment of electrical capacitance that can account simultaneously for interaction-induced correlations between charges on opposite plates and the wave nature of electrons. In contrast to the common approaches outlined above, we use linear-response theory and obtain the capacitance from suitably defined charge susceptibilities, in a way similar to what was done for Wigner crystals \cite{Giamarchi-2003-2, Chitra-2005-1}. We employ the machinery of many-body theory to express these susceptibilities in terms of charge polarizabilities that describe the effects of both intra- and interplate interactions, and show that these polarizabilities are the fundamental quantities determining the capacitance. We obtain a general expression for the quantum capacitance that reduces to the known one [derived from Eq.~(\ref{eq:CE})] in appropriate limits, but that is also valid in the presence of strong interplate interactions, in which case it contains an additional contribution that we refer to as the cross quantum capacitance. Our calculations show that the conventional and the cross quantum capacitances contribute on equal footing when the plate separation is very small. Through a perturbative analysis, we predict a non-monotonic evolution of the total capacitance upon increasing separation between the plates, whose observation would provide the ultimate experimental evidence for the presence of a cross quantum capacitance term.

We emphasize that---although the influence of interlayer correlations to the quantum capacitance has not been considered before---there is little doubt that these interactions play an important role on different physical phenomena when two electronic systems are brought into sufficiently close proximity. Coulomb drag effects in transport provide a clear example, to which considerable experimental and theoretical efforts have been devoted \cite{Narozhny-2016}. Particularly relevant for our paper here is the theoretical study of interlayer correlations in homogeneous electron-electron and electron-hole double layers \cite{Zheng-1994, [See ][ and references therein.]Liu-1996}, using the Singwi--Tosi--Land--Sj\"{o}lander (STLS) local-field theory for the dielectric function \cite{Singwi-1968, Mahan}. These studies---that mostly focused on the correlation energy, pair distribution functions and plasmon excitations---emphasize the role of screening and imply that the screening of charges on one plate by the polarization of the other plate may become as important as the intraplate screening in atomically-thin capacitors. Indeed, cross-screening effects were beautifully demonstrated in experiments on graphene/h-BN/metal structures, in which the dispersion of acoustic plasmon in graphene was found to depend on the h-BN thickness \cite{Iranzo-2018}. As we will discuss in this paper, the relevance of the conventional quantum capacitance term and the cross quantum capacitance that we introduce is directly related to the interplay of intra and interlayer screening processes.

Finally, we mention that the work presented in this paper is motivated by recent experiments on ionic liquid gated monolayer transition metal dichalchogenides \cite{Zhang-2019}, in which the evolution of the capacitance measured as a function of electron density exhibits large quantum capacitance effects that seemingly cannot be reconciled with existing theory. In these devices, the ions at the surface of the ionic liquid act as one of the capacitor plates and the charge (electrons or holes) accumulated on the monolayers form the other plate. As the ions are in direct contact with the monolayer, the separation between the two plates is only a few Angstr\"{o}ms. The geometrical capacitance is therefore enormous ($\sim50~\mu\mathrm{F/cm}^2$)---vastly facilitating the observation of quantum capacitance phenomena---and interplate interactions are as large as intraplate ones---enhancing the role of cross quantum capacitance effects. Although the use of ionic liquids for gating makes these systems difficult to model realistically \cite{Loth-2010, Ma-2015, Petach-2017, Bera-2019} (and indeed it is not the aim of this paper to model them specifically), all qualitative aspects of ionic gated transition metal dichalcogenide monolayers appear to be ideal to maximize the effect of interaction-induced interlayer correlations. That is why we believe that the physical phenomena that we are discussing here are certainly observable experimentally.

This paper is structured as follows. We define our system in Sec.~\ref{sec:CQ-Pi}, recall basic notions about the capacitance (Sec.~\ref{sec:C-E}), present the linear-response calculation (Sec.~\ref{sec:C-chi}), and discuss various implications (Sec.~\ref{sec:CQ}). These include, in particular, the failure of random phase approximation (RPA). In Sec.~\ref{sec:CQ-eps}, we discuss the relation between the quantum capacitance and electronic screening. Approximations are proposed in Sec.~\ref{sec:approx}, in particular a diagrammatic approach that goes beyond RPA. We study the cases of symmetric and asymmetric capacitors. Appendices \ref{app:3D} to \ref{app:comparison} complement the information given in the main text. A brief account of some of the results presented here has already been given as Supporting Information in Ref.~\onlinecite{Zhang-2019}.

\section{Quantum capacitance and polarizabilities}
\label{sec:CQ-Pi}

\begin{figure}[b]
\includegraphics[width=0.8\columnwidth]{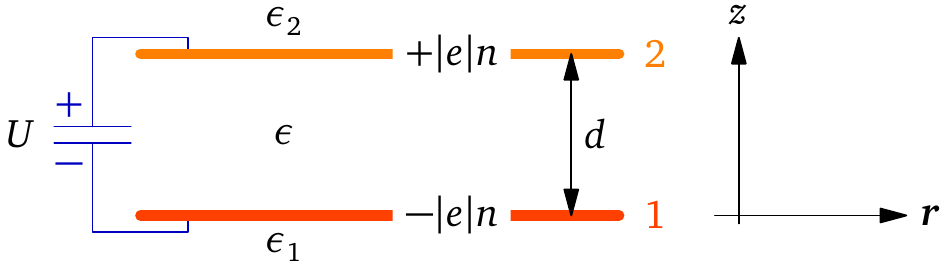}
\caption{\label{fig:capacitor}
The ideal quantum capacitor. The infinite 2D plates labeled ``1'' and ``2'' carry opposite excess charges with an average density $n$ in response to the applied voltage $U$. Quantum tunneling between the plates is forbidden.
}
\end{figure}

We consider a system composed of two infinite two-dimensional metallic plates separated by a distance $d$ (Fig.~\ref{fig:capacitor}). For simplicity, we assume in the main text a uniform background with constant permittivity $\epsilon_1=\epsilon_2=\epsilon$ (the generalization is presented in Appendix~\ref{app:epsilon}). The plates may be different with arbitrary electronic dispersions, but we shall take parabolic dispersions for illustrations. With the application of a bias voltage $U$ (we reserve the symbol $V$ for the Coulomb interaction) opposite charges are brought to the plates, maintaining overall charge neutrality. We assume that the state of the biased system is characterized by average excess charge densities of opposite signs in the two plates, $\rho_1=-|e|n$ and $\rho_2=+|e|n$. The specific capacitance $C$ (capacitance per unit area) is defined as
	\begin{equation}\label{eq:C}
		|e|n=CU
	\end{equation}
in the limit of vanishing $U$. In Sec.~\ref{sec:C-chi}, we rely on this definition to evaluate the capacitance as the linear response of the system to the applied bias.

\subsection{Capacitance and electronic energy}
\label{sec:C-E}

In order to put our approach in perspective, we discuss here the more popular definition of the linear capacitance given by Eq.~(\ref{eq:CE}) in the limit $n\to0$. Equations (\ref{eq:CE}) and (\ref{eq:C}) are equivalent \footnote{The free energy of the capacitor is $F(n)=E(n)-e\frac{U}{2}(n_1+n)+e\frac{U}{2}(n_2-n)$, where $n_1$ and $n_2$ are the equilibrium electronic densities in the plates and $E(n)$ is the total energy, not including the applied bias. As a given bias $U$, $n$ adjusts such as to minimize $F(n)$, hence $U=\frac{1}{e}\frac{dE}{dn}$. Inserting this in Eq.~(\ref{eq:C}), which is equivalent to $\frac{1}{C}=\frac{1}{e}\left.\frac{dU}{dn}\right|_{n=0}$, leads to Eq.~(\ref{eq:CE}).}, but dictate very different analytical methods for computing the capacitance. With the former, one has to model the various contributions to the energy, which is typically done by separating kinetic, exchange, and correlation terms \cite{Kopp-2009}. With the latter, one has instead to model response functions.

The Hamiltonian of the capacitor is $H_1+H_2+H_ {12}+H'$, where $H_1$ and $H_2$ collect the kinetic and Coulomb energies in both plates, $H_{12}$ is the Coulomb interaction between charges on opposite plates, and $H'$ is the energy due to the applied voltage. In the capacitor geometry, the Coulomb interaction is $|e|/(2\epsilon q)$ for two charges on the same plate and $-|e|e^{-qd}/(2\epsilon q)$ for two charges on different plates, where $\vec{q}$ is the two-dimensional wave vector (see Appendix~\ref{app:3D}). The electrostatic (Hartree) contribution to the energy per unit surface $S$ is then readily evaluated to be:
	\begin{align}\label{eq:Hartree}
		\nonumber
		E_{\mathrm{H}}&=\frac{1}{S}\int d^2rd^2r'\left[
		\frac{1}{2}\frac{\rho_1(\vec{r})\rho_1(\vec{r}')}{4\pi\epsilon|\vec{r}-\vec{r}'|}
		+\frac{1}{2}\frac{\rho_2(\vec{r})\rho_2(\vec{r}')}{4\pi\epsilon|\vec{r}-\vec{r}'|}\right.\\
		\nonumber
		&\quad\left.+\frac{\rho_1(\vec{r})\rho_2(\vec{r}')}{4\pi\epsilon\sqrt{|\vec{r}-\vec{r}'|^2+d^2}}\right]\\
		&=\lim_{q\to0}\frac{n^2e^2}{2\epsilon q}\left(\frac{1}{2}+\frac{1}{2}
		-e^{-qd}\right)=\frac{n^2e^2}{2\epsilon}d.
	\end{align}
The diverging repulsions between charges on both plates (with the factors of $1/2$ avoiding double-counting) are canceled by a diverging attraction between charges on opposite plates, leaving a finite Hartree energy and the so-called geometrical capacitance
	\begin{equation}\label{eq:CG}
		\frac{1}{\CG}=\frac{1}{C_{\mathrm{H}}}=\frac{d}{\epsilon}.
	\end{equation}
The leftover electrostatic energy can be regarded as the classical energy of the electric field ($\vec{E}=-|e|n/\epsilon\hat{\vec{z}}$, with $z$ the coordinate perpendicular to the plates) given by $\frac{1}{S}(\epsilon/2)\int d^3r|\vec{E}|^2=e^2n^2d/(2\epsilon)$. This observation emphasizes that the geometrical capacitance results from charge conservation and is not altered by the dynamical screening of the Coulomb potential inside and in-between the plates. As the capacitances add in series in Eq.~(\ref{eq:CE}), the geometrical capacitance easily dominates when the plates are separated by a large distance. Otherwise, the kinetic and interaction energies of the excess charges also contribute various terms.

In this paper, we use the denomination ``quantum capacitance'' for the sum of \emph{all} contributions but the geometrical one. In the literature, though, ``quantum capacitance'' often refers to just the bare kinetic energy \cite{Luryi-1988}, or to the quasiparticle energy \cite{Xia-2009}. For clarity, we will call ``kinetic capacitance'' the term associated with the kinetic energy. For a parabolic dispersion with mass $m$ in two dimensions, the kinetic energy at zero temperature is $\pi\hbar^2n^2/(2m)$ and the kinetic capacitance is $C_{\mathrm{kin}}=me^2/(\pi\hbar^2)$. The ``cross quantum capacitance'' effects of interest here originate from nonclassical interplate correlations. Specifically, if we collect all contributions to the energy, the total capacitance may be decomposed as
	\begin{equation}\label{eq:C-E}
		\frac{1}{C}=\frac{1}{C_{\mathrm{H}}}+\frac{1}{\Ck}+\frac{1}{C_{\mathrm{kin},2}}
		+\frac{1}{C_{\mathrm{xc},1}}+\frac{1}{C_{\mathrm{xc},2}}+\frac{1}{C_{\mathrm{xc}}}.
	\end{equation}
The $C_{\mathrm{xc},\alpha}$'s relate to the exchange \footnote{For the two-dimensional electron gas, the exchange contribution is $C_{\mathrm{x},\alpha}=-2\pi\epsilon\sqrt{2\pi n_{\alpha}}$ with $n_{\alpha}$ the equilibrium electronic density in plate $\alpha$. Hence, defining $\pi(r_sa_0)^2=1/n_{\alpha}$ with $a_0$ the Bohr radius, we have $1/C_{\mathrm{kin},\alpha}+1/C_{\mathrm{x},\alpha}=0$ for $r_s=\pi/\sqrt{2}$. See also Eq.~(17) of Ref.~\onlinecite{Tanatar-1989}.} and correlation energies of the two isolated plates, which for the homogeneous electron gas have been evaluated by quantum Monte Carlo \cite{Tanatar-1989, Drummond-2009}. The remaining term, $C_{\mathrm{xc}}$, contains everything not counted in the other terms and is due to the interplate interaction. This includes energetic contributions that can be interpreted as exchange and correlation effects among charges in different plates, but it also includes terms that can be regarded as modifications of the bare kinetic and intraplate exchange-correlation energies due to the interplate interaction. We underline that for an isolated plate the partition of the total electronic energy into kinetic and exchange-correlation terms is to some degree arbitrary---although one can always \emph{define} the various contributions by some partition---and that the situation worsens when the capacitor is formed. Therefore, Eq.~(\ref{eq:C-E}) must be understood as a definition of $C_{\mathrm{xc}}$, which shows that the difficulty of calculating this term is the same as the difficulty of calculating the exact capacitance. Since there is no obvious way to model $C_{\mathrm{xc}}$, this term is generally neglected and thus cross quantum capacitance effects are missed.

\subsection{Capacitance and charge susceptibilities}
\label{sec:C-chi}

We now move on to discuss the capacitance calculated within linear-response theory. We model the effect of an applied voltage by opposite shifts of the chemical potentials in the two plates: if plate~1 is brought to a uniform potential $-U/2$ and plate~2 to $+U/2$, this energy is \footnote{The Hamiltonian $H'$ must be invariant if, instead of the symmetric choice $U_1=-U/2$ and $U_2=+U/2$, the plates are brought to arbitrary potentials such that $U_2-U_1=U$. For this purpose, we should write in general $H'=-\Delta\mu_1N_1-\Delta\mu_2N_2+\frac{1}{2}(U_1+U_2)(-|e|N_1+|e|N_2)$, where the first two terms represent the addition of charges and the last term accounts for a global shift of all energy levels in each plate. Here, $\Delta\mu_1=-|e|U_1$, $\Delta\mu_2=+|e|U_2$, and $N_1$, $N_2$ is the total number of electrons, respectively holes, in plate~1, 2. This is equivalent to Eq.~(\ref{eq:bias}).}
	\begin{equation}\label{eq:bias}
		H'=-\frac{|e|U}{2}\left[\int d^2r_1\,\hat{n}_1(\vec{r}_1)+\int d^2r_2\,\hat{n}_2(\vec{r}_2)\right],
	\end{equation}
where $\hat{n}_{\alpha}(\vec{r}_{\alpha})$ is the number-density (not charge-density) operator in the plate $\alpha$ (our sign convention is $\hat{\rho}_1=-|e|\hat{n}_1$ and $\hat{\rho}_2=+|e|\hat{n}_2$, such that $\hat{n}_1$ is an electron density while $\hat{n}_2$ is a positive-charge density akin to a hole density). Standard linear-response theory \cite{Bruus-Flensberg} in $H'$ yields the average density induced in plate~1 as
	\begin{equation}
		\langle\delta\hat{n}_1(\vec{q})\rangle=-\frac{|e|U}{2}\left[\chi_{\hat{n}_1\hat{n}_1}(\vec{q})+
		\chi_{\hat{n}_1\hat{n}_2}(\vec{q})\right],
	\end{equation}
where $\chi_{\hat{n}_{\alpha}\hat{n}_{\beta}}(\vec{q})$ is the static susceptibility corresponding to the retarded correlation function of the operators $\hat{n}_{\alpha}(\vec{q})$ and $\hat{n}_{\beta}(-\vec{q})$. The excess charge density is $-|e|\langle\delta\hat{n}_1(\vec{q}=\vec{0})\rangle$ and the capacitance defined by Eq.~(\ref{eq:C}) follows as
	\begin{equation}\label{eq:C1}
		C=-\frac{e^2}{2}\left[\chi_{11}(\vec{0})+\chi_{12}(\vec{0})\right].
	\end{equation}
We denote simply by $\chi_{\alpha\beta}(\vec{q})$ the static susceptibilities, i.e.,
	\begin{equation}\label{eq:chinn}
		\chi_{\alpha\beta}(\vec{q})=\int_{-\infty}^{\infty}dt\,\left(\textstyle-\frac{i}{\hbar}\right)\theta(t)
		\langle[\hat{n}_{\alpha}(\vec{q},t),\hat{n}_{\beta}(-\vec{q},0)]\rangle,
	\end{equation}
where the average is taken with respect to $H_1+H_2+H_{12}$ without $H'$. Considering the density induced in plate~2 rather than plate~1, the resulting capacitance is given by
	\begin{equation}\label{eq:C2}
		C=-\frac{e^2}{2}\left[\chi_{22}(\vec{0})+\chi_{21}(\vec{0})\right].
	\end{equation}
A consistent approximation for the susceptibilities must ensure that $\chi_{11}(\vec{0})+\chi_{12}(\vec{0})=\chi_{22}(\vec{0})+\chi_{21}(\vec{0})$, meaning that precisely opposite charges are induced in both plates.

Many-body theory provides a Dyson equation for the single-particle Green's function, whereby the interaction effects are incorporated in the self-energy \cite{Mahan, Bruus-Flensberg}. Similarly, the theory of the charge response and screening expresses the susceptibility $\chi$ in terms of another function, the polarizability $\Pi$. The relation between $\chi$ and $\Pi$ is
	\begin{equation}\label{eq:chi-diag}
		\chi=-\Pi-\Pi\, V\chi,
	\end{equation}
where $V$ is the Coulomb interaction. The distinction between susceptibility and polarizability may be understood as a separation between classical electrostatic effects and quantum mechanical effects, a notion that we elaborate in Appendix~\ref{app:chiclass}. This separation is crucial in the problem of the quantum capacitor, as it allows us to split the many-body formula for the capacitance into geometrical and quantum terms.

For this purpose, Eq.~(\ref{eq:chi-diag}) must be generalized to the specific geometry of the capacitor, taking into account the three-dimensional nature of the problem with a non-local screening along the $z$ direction. This is presented in Appendix~\ref{app:3D} and leads to the same relation as (\ref{eq:chi-diag}), now between the $2\times2$ matrices $\chi$ defined in Eq.~(\ref{eq:chinn}), the matrix $V$ with elements $V_{11}=V_{22}=e^2/(2\epsilon q)$ and $V_{12}=V_{21}=-e^2e^{-qd}/(2\epsilon q)$, and the matrix $\Pi_{\alpha\beta}$ giving the polarizability tensor. Since diagrammatic perturbation theory provides a recipe for evaluating directly the polarizabilities $\Pi_{\alpha\beta}(\vec{q})$ as the sum of irreducible diagrams in the expansion of $\chi_{\alpha\beta}(\vec{q})$, Eq.~(\ref{eq:chi-diag}) is a viable alternative to Eq.~(\ref{eq:chinn}) for computing the susceptibilities (Appendix~\ref{app:diagrams}).

Solving Eq.~(\ref{eq:chi-diag}) for $\chi$, we deduce the combinations of susceptibilities entering the formula for the capacitance:
\begin{widetext}
	\begin{subequations}\label{eq:chi+chi}
	\begin{align}\label{eq:chi11+chi12}
		\chi_{11}+\chi_{12}&=\frac{-\Pi_{11}-\Pi_{12}-(V_{22}-V_{12})(\Pi_{11}\Pi_{22}-\Pi_{12}\Pi_{21})}
		{1+V_{11}\Pi_{11}+V_{22}\Pi_{22}+V_{12}\Pi_{21}+V_{21}\Pi_{12}+(V_{11}V_{22}-V_{12}V_{21})
		(\Pi_{11}\Pi_{22}-\Pi_{12}\Pi_{21})}\\
		\label{eq:chi22+chi21}
		\chi_{22}+\chi_{21}&=\frac{-\Pi_{22}-\Pi_{21}-(V_{11}-V_{21})(\Pi_{11}\Pi_{22}-\Pi_{12}\Pi_{21})}
		{1+V_{11}\Pi_{11}+V_{22}\Pi_{22}+V_{12}\Pi_{21}+V_{21}\Pi_{12}+(V_{11}V_{22}-V_{12}V_{21})
		(\Pi_{11}\Pi_{22}-\Pi_{12}\Pi_{21})}.
	\end{align}\end{subequations}
\end{widetext}
The expressions (\ref{eq:chi11+chi12}) and (\ref{eq:chi22+chi21}) are equal in the limit $q\to0$ (the polarizabilities are finite at $q=0$), which proves the consistency of Eqs.~(\ref{eq:C1}) and (\ref{eq:C2}). Note that the relation $\chi_{11}+\chi_{12}=\chi_{22}+\chi_{21}$ in the limit $q\to0$ does not imply or require $\Pi_{11}+\Pi_{12}=\Pi_{22}+\Pi_{21}$ in the same limit.

Equation~(\ref{eq:chi+chi}) can also be expressed in terms of the screened Coulomb potentials. The usual many-body equation for the screened potential $W$ has the form \cite{Bruus-Flensberg}
	\begin{equation}\label{eq:W-diag}
		W=V-V\Pi W
	\end{equation}
and is confirmed for the capacitor geometry in Appendix~\ref{app:3D}. Eliminating the singular matrix elements of $V$ between Eqs.~(\ref{eq:chi-diag}) and (\ref{eq:W-diag}), we find representations of the susceptibilities in which every quantity is regular in the long-wavelength limit:
	\begin{subequations}\label{eq:chi+chi-W}\begin{align}
		\label{eq:ch11+chi12-W}\nonumber
		\chi_{11}+\chi_{12}&=-(\Pi_{11}+\Pi_{12})(1-W_{11}\Pi_{11}-W_{21}\Pi_{12})\\
		&\quad+(\Pi_{22}+\Pi_{21})(W_{12}\Pi_{11}+W_{22}\Pi_{12})\\
		\label{eq:ch22+chi21-W}\nonumber
		\chi_{22}+\chi_{21}&=-(\Pi_{22}+\Pi_{21})(1-W_{22}\Pi_{22}-W_{12}\Pi_{21})\\
		&\quad+(\Pi_{11}+\Pi_{12})(W_{21}\Pi_{22}+W_{11}\Pi_{21}).
	\end{align}\end{subequations}
We solve Eq.~(\ref{eq:W-diag}) for $W$, insert the bare interaction $V$, and take the limit $q\to0$ to get
	\begin{equation}\label{eq:W0}
		W(\vec{0})=\frac{
		\begin{bmatrix}1+e^2\Pi_{22}\frac{d}{\epsilon}&-\left(1+e^2\Pi_{12}\frac{d}{\epsilon}\right)\\
		-\left(1+e^2\Pi_{21}\frac{d}{\epsilon}\right)&1+e^2\Pi_{11}\frac{d}{\epsilon}\end{bmatrix}}
		{\Pi_{11}+\Pi_{22}-\Pi_{12}-\Pi_{21}
		+e^2\left(\Pi_{11}\Pi_{22}-\Pi_{12}\Pi_{21}\right)\frac{d}{\epsilon}},
	\end{equation}
where the polarizabilities are evaluated at $q=0$. This representation of the screened Coulomb potential will be used in Sec.~\ref{sec:dQ} to define screening lengths. Substitution of this formula into Eqs.~(\ref{eq:chi+chi-W}) shows that both expressions are indeed equal and yields a general formula for the linear capacitance:
	\begin{equation}\label{eq:C-Pi}
		\frac{1}{C}=\frac{1}{\CG}+\frac{1}{e^2}\frac{\Pi_{11}+\Pi_{22}-\Pi_{12}-\Pi_{21}}
		{\Pi_{11}\Pi_{22}-\Pi_{12}\Pi_{21}}.
	\end{equation}
This is our main result. We show in Appendix~\ref{app:epsilon} that it holds unchanged in the general case $\epsilon_1\neq\epsilon\neq\epsilon_2$.

Equation~(\ref{eq:C-Pi}) shows that the total capacitance $C$ is given by the in-series addition of the conventional geometrical capacitance (the first term on the right-hand side) and of one additional term that is the quantum capacitance
	\begin{equation}\label{eq:CQ}
		\frac{1}{\CQ}=\frac{1}{e^2}\frac{\Pi_{11}+\Pi_{22}}{\Pi_{11}\Pi_{22}-\Pi_{12}\Pi_{21}}
		-\frac{1}{e^2}\frac{\Pi_{12}+\Pi_{21}}{\Pi_{11}\Pi_{22}-\Pi_{12}\Pi_{21}}.
	\end{equation}
The quantum capacitance depends exclusively on the intra- and the interlayer polarizabilities. It differs from the conventional expression, which does not include the effect of interlayer correlations. The difference is what we refer to as cross quantum capacitance. We can separate the contributions of the conventional quantum capacitance and of the cross quantum capacitance in different ways, for instance as shown by the two terms in Eq.~(\ref{eq:CQ}). This illustrates explicitly how the cross quantum capacitance gives an additional contribution that adds in series to the conventional quantum capacitance, and that vanishes when interlayer correlations are ignored (the first term on the right-hand side of Eq.~(\ref{eq:CQ}) does indeed reduce to the known expression for the quantum capacitance when $\Pi_{12}=\Pi_{21}=0$). However, separating the quantum capacitance contributions as in Eq.~(\ref{eq:CQ}) can be misleading in some cases, because interlayer correlations can also influence the intralayer polarizabilities $\Pi_{11}$ and $\Pi_{22}$.

For this reason, in the rest of the paper we analyze the total capacitance and the quantum capacitance [as given by Eq.~(\ref{eq:C-Pi})], without singling out the cross quantum capacitance contribution. We will refer to cross quantum capacitance effects whenever we discuss phenomena that originate from interaction-induced interlayer correlations. We will first analyze the equation formally in different cases (Sec.~\ref{sec:CQ}), and then discuss different model calculations providing explicit expressions for the polarizabilities and hence for the quantum capacitance (Sec.~\ref{sec:approx}). Our main goal is to gain physical intuition about the cross quantum capacitance effects described by Eq.~(\ref{eq:C-Pi}). Key questions concern the magnitude and the sign of the interlayer polarizabilities: if $\Pi_{12}\approx\Pi_{11}$ and/or $\Pi_{21}\approx\Pi_{22}$, large quantitative effects can be expected, increasing or reducing the quantum capacitance relative to the case in which interlayer correlations are disregarded, depending on the sign.

\subsection{Quantum and cross quantum capacitances}
\label{sec:CQ}

To start gaining intuition, in this section we discuss some implications and various limits of Eq.~(\ref{eq:C-Pi}), starting with the RPA, which allows us to recover the known results. We then study how weak interplate effects correct the quantum capacitance, point out that for symmetric capacitors, the intra- and interplate polarizabilities must be treated on the same footing, and argue that the quantum capacitance must be discontinuous at a transition where the total capacitance is equal to the geometrical one.

\subsubsection{Random-phase approximation and kinetic capacitance}
\label{sec:RPA}

At the RPA level, all Coulomb effects but the classical Hartree terms responsible for $\CG$ are ignored. The intraplate polarizabilities $\Pi_{\alpha\alpha}$ reduce (at zero temperature) to the Fermi-level DOS $\nu_{\alpha}$ in each plate and the interplate polarizabilities vanish identically (Appendix~\ref{app:diagrams}). Inserting $\Pi^0_{\alpha\beta}=\delta_{\alpha\beta}\nu_{\alpha}$ in Eq.~(\ref{eq:C-Pi}) yields
	\begin{equation}\label{eq:C0}
		\frac{1}{C_0}=\frac{1}{G_{\mathrm{G}}}+\frac{1}{e^2\nu_1}+\frac{1}{e^2\nu_2}.
	\end{equation}
Hence the linear-response theory applied with the RPA susceptibilities reproduces the least accurate of the approximations based on the electronic energy for the capacitance, which only counts the unrenormalized kinetic energies and the classical Hartree term. We conclude that the RPA confirms the ability of Eq.~(\ref{eq:C-Pi}) to reproduce known results, but is too crude to capture any of the interplate phenomena giving rise to cross quantum capacitance effects.

\subsubsection{Corrections beyond RPA}

If the interplate polarizabilities are small compared to the intraplate ones, as expected for instance when the distance between the capacitor plates is sufficiently large, the quantum capacitance may be expanded as
	\begin{equation}\label{eq:first-order}
		\frac{1}{\CQ}\approx\frac{1}{e^2}\left(\frac{1}{\Pi_{11}}+\frac{1}{\Pi_{22}}\right)
		\left(1-\frac{\Pi_{12}+\Pi_{21}}{\Pi_{11}+\Pi_{22}}\right).
	\end{equation}
This expression results from a Taylor expansion of Eq.(\ref{eq:C-Pi}), but it can also be obtained by writing $1/\CQ=1/C_{\mathrm{Q},1}(1+C_{\mathrm{Q},1}/C_{\mathrm{Q},2})$, where $1/C_{\mathrm{Q},1}$ and $1/C_{\mathrm{Q},2}$ are the two terms in the right-hand side of Eq.(\ref{eq:CQ}), and then neglecting $\Pi_{12}\Pi_{21}$ relative to $\Pi_{11}\Pi_{22}$ in the expression of $C_{\mathrm{Q},1}$. The first factor corresponds to a renormalized RPA result, where the noninteracting polarizabilities $\nu_{\alpha}$ are replaced by the fully interacting polarizabilities $\Pi_{\alpha\alpha}$. The second factor is the correction due to interplate correlations. This correction enhances (reduces) the quantum capacitance if the interplate polarizabilities are positive (negative). The approximate expressions derived in Sec.~\ref{sec:approx} show that both cases are possible. Equation~(\ref{eq:first-order}) furthermore shows that cross quantum capacitance effects can be understood as a multiplicative correction to the conventional quantum capacitance, an interesting observation given the analysis of the experimental result presented in Ref.~\onlinecite{Zhang-2019}.

The expression of the quantum capacitance is particularly simple if the two plates are identical,
	\begin{equation}\label{eq:C-same}
		\CQ=\frac{e^2}{2}\left(\Pi_{11}+\Pi_{12}\right)\qquad\text{(identical plates)},
	\end{equation}
an exact formula that does not require $\Pi_{12}$ to be small. The first term may be interpreted as the kinetic capacitance renormalized by intra- (as well as inter-) plate interactions. Indeed, the polarizabilities entering Eq.~(\ref{eq:C-Pi}) are properties of the \emph{capacitor}, such that $\Pi_{\alpha\alpha}$ is in general different from the polarizability of the isolated plate $\alpha$, owing to the correlated motion of the charges in both plates. The second term in Eq.~(\ref{eq:C-same}) describes the cross quantum capacitance effect. This expression makes it very clear that if $\Pi_{12}$ becomes comparable to $\Pi_{11}$, cross-quantum capacitance effects become very large.

\subsubsection{Strong interplate effects, phase transitions}
\label{sec:W}

As the intraplate and interplate polarizabilities contribute on equal footings to $\CQ$, one can expect large cross quantum effects for ultrathin capacitors, since in this regime the interplate Coulomb correlations become important \cite{Zheng-1994, Liu-1996}, such that $\Pi_{11}$ and $\Pi_{12}$ must have similar magnitudes. This means that an interpretation of capacitance data only in terms of intraplate DOS or thermodynamic compressibility is bound to fail.

A value $\Pi_{12}\approx\Pi_{11}$ in Eq.~(\ref{eq:C-same}) means a factor two increase of $\CQ$ relative to its value in the absence of interplate correlations. This is certainly not a small effect---in systems like batteries, a factor of two increase in the charge stored \emph{is} a major step---but it remains nevertheless an effect of quantitative rather than qualitative nature. On the other hand, a qualitative change must occur when $\Pi_{12}$ is large and negative, approaching $-\Pi_{11}$, such that $\CQ$ approaches zero. In fact, because the thermodynamic stability of the capacitor requires that the total capacitance $C=(1/\CG+1/\CQ)^{-1}$ is positive, $\CQ$ is not allowed to continuously cross zero: It is forbidden to take values in the range extending from zero to $-\CG$. Either $\CQ$ is positive, in which case $C<\CG$, or negative and smaller than $-\CG$, in which case $C>\CG$. We can distinguish two kinds of transitions from a regime where $C<\CG$ to a regime where $C>\CG$, depending on whether the total capacitance is continuous or not at the transition.

A discontinuous jump of $C$ at $C=\CG$ is associated with a \emph{finite} discontinuity of $\CQ$ at the transition. This behavior was found in a numerical Hartree--Fock study of electron-hole bilayers with Dirac-like dispersion in a strong magnetic field \cite{Roostaei-2015}. This phenomenology was not observed, however, in the experiments that have reported $C>\CG$  \cite{Eisenstein-1992, Li-2011}. In those experiments, where the control parameter is the electronic density, $C$ crosses smoothly from values $C<\CG$ at high density to values $C>\CG$ at low density. This means that $\CQ$ has an \emph{infinite} discontinuity at the transition, jumping from $+\infty$ on the high-density side to $-\infty$ on the low-density side. In the context of the one-plate approaches, where the second plate is treated classically, $C=\CG$ signals a divergence and sign change of the charge susceptibility, a behavior that is normally associated with an electronic phase transition. Indeed, in the regime where this phenomenon is observed, a two-dimensional electron gas is expected to undergo various magnetic and localization phase transitions with decreasing density \cite{Drummond-2009}.

Within many-body theory, the occurrence of an ordering phase transition coincides with a divergence of the susceptibility associated with the observable that orders. A charge-density wave (CDW), for instance, is a static charge order that implies a divergence of the charge susceptibility $\chi(Q)$ at the ordering wave-vector $Q$. In bulk systems, this happens whenever $1+\Pi(Q)V(Q)=0$ [see Eq.~(\ref{eq:chi-diag})]. Fulfilling this condition does not require a singular behavior of $\Pi(q=0)$. It follows that in a direct transition from the homogeneous paramagnetic state into a CDW state like a Wigner crystal---where calculations indeed suggest that $C>\CG$ \cite{Skinner-2010, Chitra-2005-1}---we do not necessarily expect the long-wavelength polarizability to diverge and, therefore, we do not expect the quantum capacitance to diverge either. Such a transition without divergence of $\CQ$ would be characterized by a discontinuity of $C$. On the contrary, we expect a capacitance continuously crossing the value $C=\CG$ at a transition to a homogeneous magnetic phase. Such a transition is characterized by a diverging spin susceptibility at $q=0$ and in the paramagnetic state, the spin susceptibility is proportional to the polarizability, such that we expect the quantum capacitance to diverge in that case.

The above considerations valid for bulk systems illustrate the concept, but the capacitor geometry offers additional possibilities, for instance a divergence of $\Pi_{12}(q=0)$ associated with the formation of a liquid phase of interplate bound pairs. Preliminary works suggest that the ground-state phase diagram of the ideal capacitor may be rich \cite{Hanna-2000}. Thus the exact sequence of phase transitions as a function of decreasing density in the ultrathin capacitor may hold surprises, especially if the equilibrium densities in the two plates are incommensurate. More work is needed in order to explore these questions, starting from a generalization of Eq.~(\ref{eq:C-Pi}) to situations where the charges in the plates break spin-rotation symmetry.

\subsection{Quantum capacitance and electronic screening}
\label{sec:CQ-eps}

The polarizability is closely related to the phenomenon of screening, as illustrated for instance in Eq.~(\ref{eq:W0}). We explore this connection further in this section. First, we relate the sign of the interplate polarizabilities to the behavior of screening charges in the capacitor. In a second part, we introduce a number of screening lengths and relate them to the quantum capacitance.

\subsubsection{Interplate polarizability and screening charges}
\label{sec:screening}

The different contributions to the capacitance add up in series [see, e.g., Eq.~(\ref{eq:C-E})]. Here, we discuss an alternative point-of-view, where one adds up charges rather than inverse capacitances. This allows us to split the capacitor response into ``external'' charges and screening charges, and furthermore to distinguish intraplate from interplate screening charges. In this way, we can relate the sign of the interplate polarizability to the sign of interplate screening charges.

\begin{figure*}[t]
\includegraphics[width=0.9\textwidth]{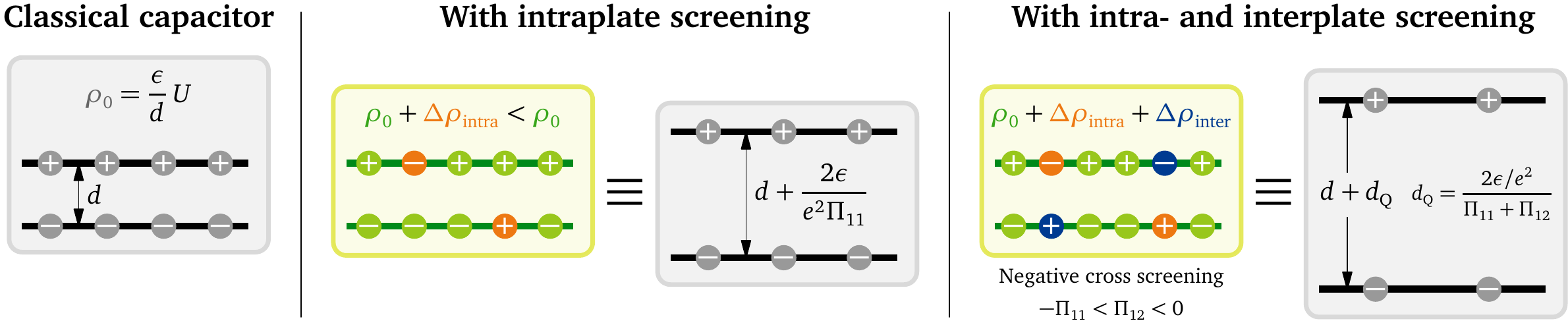}
\caption{\label{fig:dQ}
According to classical electrostatics, a potential difference $U$ applied across the capacitor yields an excess charge $(\epsilon/d)U$ on the plates (left). If the plates have finite polarizability, screening charges oppose to the charges induced by the potential difference and the net amount of excess charge is less for the same $U$: This is equivalent to a classical capacitor of increased thickness (middle). The ``negative'' cross screening contributes screening charges that have the same sign as the intraplate screening charges: this further reduces the net excess charge, increases the quantum capacitance length, and reduces the quantum capacitance.
}
\end{figure*}

We express the total excess density on the plates as $n=n_{\mathrm{G}}-n_{\mathrm{intra}}-n_{\mathrm{inter}}$. $n_{\mathrm{G}}$ represents the charge one would have if all quantum capacitance effects were absent, i.e., $|e|n_{\mathrm{G}}=\CG U$. This charge is reduced by intraplate screening, which means that a density $n_{\mathrm{intra}}$ of opposite-sign charges is attracted to the plate in response to the ``external'' charge $|e|n_{\mathrm{G}}$. We define $n_{\mathrm{intra}}$ such that $n_{\mathrm{G}}-n_{\mathrm{intra}}$ gives the total charge one would have if interplate quantum effects were absent ($\Pi_{12}=\Pi_{21}=0$). Finally, the capacitor charge is further modified by a density $n_{\mathrm{inter}}$ of screening charges displaced by interplate correlations. Equation~(\ref{eq:C}) shows that $n/n_{\mathrm{G}}=C/\CG$, which, together with Eq.~(\ref{eq:C-Pi}) and the definitions of $n_{\mathrm{intra}}$ and $n_{\mathrm{inter}}$, gives
	\begin{subequations}\label{eq:screeniong-charges}\begin{align}
		\frac{n_{\mathrm{intra}}}{n_{\mathrm{G}}}&=\frac{\frac{1}{\scriptstyle\Pi_{11}}+\frac{1}{\Pi_{22}}}
		{\scriptstyle\frac{e^2d}{\epsilon}+\frac{1}{\Pi_{11}}+\frac{1}{\Pi_{22}}}\\
		\frac{n_{\mathrm{inter}}}{n_{\mathrm{G}}}&=\frac{\scriptstyle-\Pi_{12}-\Pi_{21}+\Pi_{12}\Pi_{21}
		\left(\frac{1}{\Pi_{11}}+\frac{1}{\Pi_{22}}\right)}
		{\scriptstyle\left(\frac{e^2d}{\epsilon}+\frac{1}{\Pi_{11}}+\frac{1}{\Pi_{22}}\right)
		\left[\frac{\epsilon}{e^2d}(\Pi_{11}+\Pi_{22}-\Pi_{12}-\Pi_{21})
		+\Pi_{11}\Pi_{22}-\Pi_{12}\Pi_{21}\right]}.
	\end{align}\end{subequations}
$n_{\mathrm{intra}}$ drops from the value $n_{\mathrm{G}}$ when the polarizabilities are small to zero when they are large. As discussed in Appendix~\ref{app:chiclass}, the limit $\Pi\to\infty$ corresponds to the limit of classical electrostatics, where indeed $n_{\mathrm{intra}}=0$ and $n=n_{\mathrm{G}}$. Depending on the sign of the interplate polarizabilities, $n_{\mathrm{inter}}$ can be positive or negative. In the following, we will use for brevity the expressions ``positive cross screening'' and ``negative cross screening''. In these mnemonics, ``positive'' and ``negative'' refer to the effect that the screening has on the total capacitance---a positive/negative cross screening increases/decreases the total capacitance. Microscopically, a positive cross screening counteracts the intraplate screening ($n_{\mathrm{inter}}<0$), thus enhancing the final density and raising the total capacitance. On the contrary, a negative cross screening reinforces the intraplate screening and reduces the density, thus lowering the total capacitance (see also Fig.~\ref{fig:dQ}). In the regime of weak interplate effects ($\Pi_{12}\ll\Pi_{11}$, $\Pi_{21}\ll\Pi_{22}$), we have
	\begin{equation*}
		\frac{n_{\mathrm{inter}}}{n_{\mathrm{G}}}=-\frac{\Pi_{12}+\Pi_{21}}{\Pi_{11}\Pi_{22}}
		\frac{\frac{e^2d}{\epsilon}}{\left(\frac{e^2d}{\epsilon}+\frac{1}{\Pi_{11}}+\frac{1}{\Pi_{22}}\right)^2}
		+\mathcal{O}(\Pi_{12}\Pi_{21}).
	\end{equation*}
Hence the ``positive'' screening behavior is characterized---with our sign convention---by a positive total interplate polarizability. This is consistent with Eq.~(\ref{eq:C-same}), where a positive $\Pi_{12}$ enhances the quantum capacitance. The study of Eq.~(\ref{eq:screeniong-charges}) for a symmetric capacitor confirms that the sign of $n_{\mathrm{inter}}$ is opposite to the sign of $\Pi_{12}$ and reveals that its magnitude can be a sizable fraction of $n_{\mathrm{intra}}$ if $\Pi_{12}$ is comparable to $\Pi_{11}$.

\subsubsection{Quantum capacitance and screening lengths}
\label{sec:dQ}

As the phenomenon of dielectric screening is often characterized by a screening length, we provide here a slightly different perspective on the quantum capacitance in terms of these lengths. This representation will be used in Sec.~\ref{sec:symmetric}, in order to build a quantum capacitance model that takes the screening length as one of its parameters.

By analogy with the geometrical capacitance, it is convenient to characterize the quantum capacitance by a length defined such that
	\begin{equation}\label{eq:C-dQ}
		\frac{1}{C}=\frac{d}{\epsilon}+\frac{\dQ}{\epsilon}.
	\end{equation}
$\dQ$ has been coined \emph{quantum capacitance length} \cite{Skinner-2013-2, Roostaei-2015}. Its expression in terms of the macroscopic polarizabilities is:
	\begin{equation}\label{eq:delta}
		\dQ=\frac{\epsilon}{e^2}\frac{\Pi_{11}+\Pi_{22}-\Pi_{12}-\Pi_{21}}
		{\Pi_{11}\Pi_{22}-\Pi_{12}\Pi_{21}}.
	\end{equation}
In the RPA ($\Pi_{\alpha\beta}\to\Pi^0_{\alpha\beta}=\delta_{\alpha\beta}\nu_{\alpha}$), the length $\dQ$ becomes $\dQ^0=(\ell^{\mathrm{TF}}_1+\ell^{\mathrm{TF}}_2)/2$, where the Thomas--Fermi lengths are defined in the usual way (for two-dimensional systems):
	\begin{equation}\label{eq:lTF}
		\frac{1}{\ell^{\mathrm{TF}}_{\alpha}}=\frac{e^2}{2\epsilon}\nu_{\alpha}.
	\end{equation}
Equation~(\ref{eq:C-dQ}) may also be recast as
	\begin{equation}\label{eq:C-dQ-TF}
		\frac{1}{C}=\frac{1}{\CG}+\frac{1}{\Ck}+\frac{1}{C_{\mathrm{kin},2}}+
		\frac{\dQ-\frac{1}{2}\left(\ell^{\mathrm{TF}}_1+\ell^{\mathrm{TF}}_2\right)}{\epsilon}.
	\end{equation}
We have pulled out the kinetic capacitances proportional to the bare DOS, $C_{\mathrm{kin},\alpha}=e^2\nu_{\alpha}$, and corrected with the Thomas--Fermi lengths defined in Eq.~(\ref{eq:lTF}). This form provides a connection with Eq.~(\ref{eq:C-E}). Apart from $\CG$, the linear-response approach does not lead to a splitting of the capacitance in terms that can be associated with distinct contributions to the electronic energy. Equation~(\ref{eq:C-dQ-TF}) nevertheless shows that the exchange-correlation terms may be interpreted as a correction to the quantum capacitance length with respect to the average of the Thomas--Fermi lengths. As it misses these corrections in the long-wavelength limit, the RPA predicts a capacitance lacking exchange and correlation effects.

The Thomas--Fermi lengths characterize the behavior of the screened Coulomb potential when the screening is treated in the Thomas--Fermi approximation. In tree dimensions, the bare potential $\sim1/q^2$ becomes $1/(q^2+\ell_{\mathrm{TF}}^{-2})$, corresponding in real space to a short-range Yukawa potential $\sim e^{-r/\ell_{\mathrm{TF}}}/r$. In two dimensions, $V(q)=e^2/(2\epsilon q)$ becomes $W(q)=e^2/[2\epsilon(q+\ell_{\mathrm{TF}}^{-1})]$ and in real space $W(r)$ crosses over from $1/r$ to $1/r^3$ at the distance $\ell_{\mathrm{TF}}$ \cite{Stern-1967-2, Giuliani-Vignale}. Hence $\ell_{\mathrm{TF}}$ measures both the strength---i.e., the spatial average $W(q=0)$---and the range---the behavior at large distance---of the screened potential. For more general screening models, it may be necessary to consider two ``screening lengths'': one for the strength ($q=0$) and one for the range ($r\to\infty$). In the following, we relate the exact quantum capacitance length $\dQ$ to screening lengths of the former kind, defined from the $q=0$ value of the Coulomb potential screened by the capacitor, Eq.~(\ref{eq:W0}).

Each matrix element of the screened Coulomb potential is parametrized by a length according to
	\begin{equation}\label{eq:lalphabeta}
		\ell_{\alpha\beta}=\frac{2\epsilon}{e^2}W_{\alpha\beta}(\vec{0}).
	\end{equation}
The intraplate lengths $\ell_{\alpha\alpha}$ belong to the capacitor---they depend on $d$---but they approach properties of the isolated plates when $d$ increases. Taking the limit $d\to\infty$ in Eq.~(\ref{eq:W0}), one finds that $\ell_{\alpha\alpha}$ approaches
	\begin{equation}\label{eq:lalpha}
		\frac{1}{\ell_{\alpha}}\equiv\lim_{d\to\infty}\frac{1}{\ell_{\alpha\alpha}}=
		\frac{e^2}{2\epsilon}\lim_{d\to\infty}\Pi_{\alpha\alpha}.
	\end{equation}
This is the generalization of Eq.~(\ref{eq:lTF}), which uses the exact polarizability of the isolated plate rather than the RPA value. Eliminating the polarizabilities from Eq.~(\ref{eq:delta}) in favor of the lengths $\ell_{\alpha\beta}$ by means of Eqs.~(\ref{eq:W0}) and (\ref{eq:lalphabeta}), we arrive at
	\begin{equation}\label{eq:delta-ell}
		\dQ=\frac{\ell}{1-\ell/d},\qquad
		\ell=\frac{1}{2}\left(\ell_{11}+\ell_{22}+\ell_{12}+\ell_{21}\right).
	\end{equation}
Note that $\ell_{\alpha\beta}$ has the sign of $W_{\alpha\beta}$, such that $\ell_{12}$ and $\ell_{21}$ are negative in the usual conditions of an attractive interplate interaction. Combined with Eqs.~(\ref{eq:C-dQ}) and (\ref{eq:CG}), Eq.~(\ref{eq:delta-ell}) leads to the remarkably simple, yet exact result
	\begin{equation}
		C=\CG\left(1-\ell/d\right).
	\end{equation}
This relation proves that the quantum correction to the geometrical capacitance is entirely determined by the strength of the screened Coulomb potential.

Because $\ell$ is a combination of screening lengths with different signs, its interpretation is not straightforward. We just mention a few limiting cases. Thermodynamic stability ($C>0$) and the assumption that $\CQ>0$ ($\dQ>0$) constrain the values of $\ell$ to the interval $0<\ell<d$. The limit of classical electrostatics (infinite polarizabilities, see Appendix~\ref{app:chiclass}) implies $W(\vec{0})=0$, in which case all $\ell_{\alpha\beta}$ vanish and $\ell$ approaches zero. The RPA gives $\ell^{\mathrm{RPA}}=(\ell_1^{\mathrm{TF}}+\ell_2^{\mathrm{TF}})/[2+(\ell_1^{\mathrm{TF}}+\ell_2^{\mathrm{TF}})/d]$, which is consistent with Eq.~(\ref{eq:C0}). For identical plates, one finds the expression $\ell=d/[1+e^2(\Pi_{11}+\Pi_{12})d/(2\epsilon)]$, which shows that $\ell$ approaches $d$ (vanishing total capacitance) when $\Pi_{12}$ approaches $-\Pi_{11}$, consistently with Eq.~(\ref{eq:C-same}).

\section{Models and approximations}
\label{sec:approx}

Evaluating exactly the interplate polarizabilities falls in the same class of difficulty as solving the problem of interacting fermions in two dimensions, which as of now requires a numerical approach like quantum Monte Carlo (QMC). While this calculation has been performed for isolated two-dimensional electron gases \cite{Tanatar-1989, Drummond-2009}, the case of two gases coupled by the long-range Coulomb interaction is challenging and has not been treated so far. The STLS method---easier than QMC but not exact---would allow one to gain useful insights. This approach was applied to bilayer systems in the past \cite{Zheng-1994, [See ][ and references therein.]Liu-1996}, but not to calculate the quantum capacitance.

Here we propose two approximations that do not rely on numerics. They allow us to predict general trends that will have to be checked by future exact calculations. Directly modeling $\Pi_{12}$ and $\Pi_{21}$ is perilous, because this quantity has no classical analog and intuition is lacking. Note that the interplate exchange interaction---which in general contributes to $\Pi_{12}$ and can be treated exactly at the Hartree--Fock level---requires overlap of the wave functions centered on opposite plates and is therefore absent for the system of Fig.~\ref{fig:capacitor}, where tunneling is forbidden. Consequently, the physics contained in $\Pi_{12}$ involves genuine correlations beyond exchange.

The first approximation applies to capacitors with identical plates in a regime where interplate correlations are weak. We express $\Pi_{12}$ in terms of a screening length, for which intuition is available and modeling is possible. We find that $\Pi_{12}$ is positive in that case, and that the quantum capacitance increases upon reducing the equilibrium density in the plates.

The second approximation targets capacitors with dissimilar plates, i.e., hosting electrons with different masses at different equilibrium densities. We use the many-body diagrammatic perturbation theory and evaluate the important diagrams approximately. When the limit of identical plates is taken, this approximation confirms that $\Pi_{12}$ is positive and agrees qualitatively with the first approximation. Otherwise, it shows that $\Pi_{12}$ is generally negative, reducing the quantum capacitance. There is a competition between two antagonistic mechanisms, which in certain parameter regimes leads to a non-monotonic dependence of the cross quantum capacitance on capacitor thickness, with the surprising consequence that the total capacitance may actually grow with making the capacitor thicker.

\subsection{Symmetric capacitor}
\label{sec:symmetric}

In order to get a first hint at the sign of $\Pi_{12}$, we consider the simplest setup, in which the plates of the capacitor in Fig.~\ref{fig:capacitor} are two ideal (disorder free) and identical electron gases. This system has only two parameters that define two fundamental length scales, the interparticle distance $d_1$ in the plates, related to the equilibrium density through $\pi d_1^2=1/n_1$, and the interplate distance $d$. A third length scale is the effective two-dimensional Bohr radius $a_1=2\pi\epsilon\hbar^2/(m_1e^2)$, which sets the strength of the Coulomb interaction.

\begin{figure}[tb]
\includegraphics[width=0.8\columnwidth]{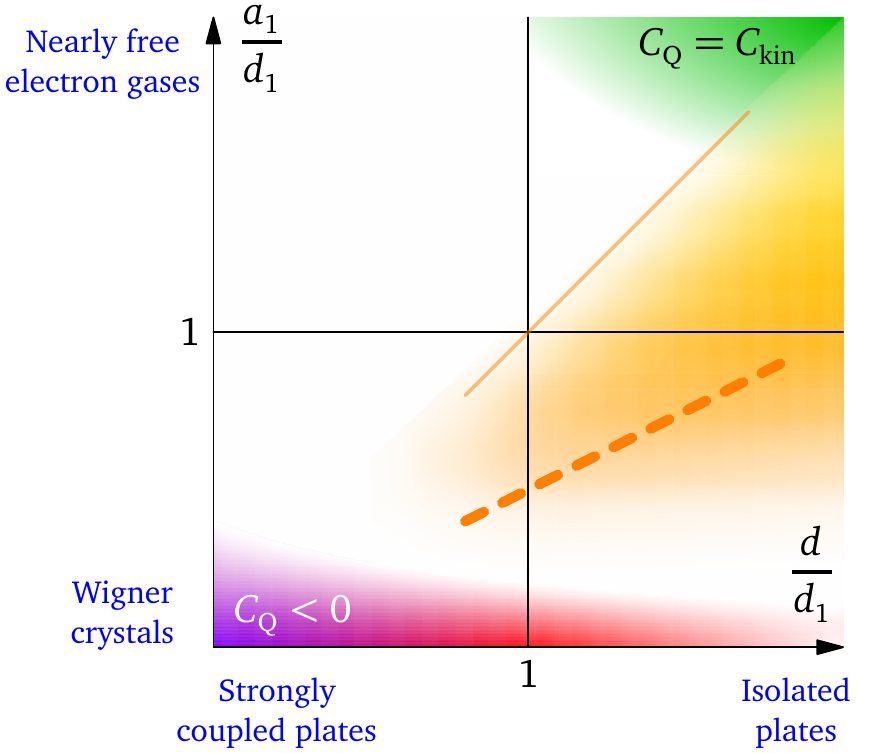}
\caption{\label{fig:regimes}
Parameter regimes for the ideal symmetric capacitor, with $a_1$ the effective Bohr radius, $d_1$ the interparticle distance, and $d$ the interplate separation. Green indicates the RPA regime, purple the localized-dipole regime, and orange the domain of validity of Eq.~(\ref{eq:model1}), which is limited by the condition $\Ck>\CG$ ($d>a_1$). The dashed line shows the direction of the horizontal axis in Fig.~\ref{fig:identical-plates}.
}
\end{figure}

If $d_1\gg a_1$, the electrons form at low temperature Wigner crystals in the plates. Since the densities are commensurate, the two Wigner crystals align in a staggered fashion in order to minimize the energy \cite{Goldoni-1996}. If furthermore $d\ll d_1$ (purple region in Fig.~\ref{fig:regimes}), the extra opposite charges induced in the plates by a finite bias are expected to bind and form an ordered array of dipoles with $1/r^3$ repulsion. The electrostatic energy of such an array is lower than that of an equivalent density of delocalized charges, and correspondingly the capacitance exceeds the geometrical capacitance, with a dependence on density estimated to be $C/\CG=(0.76\,d/d_1)^{-1}$ \cite{Skinner-2010}. The general formula Eqs.~(\ref{eq:C1}) and (\ref{eq:C-Pi}) remain applicable in this case, although modeling the uniform susceptibilities or polarizabilities is difficult. The approximations that we will present in this section assume uniform densities on the plates and cannot reach a highly inhomogeneous regime such as a Wigner crystal.

In the opposite limit of high density and weakly coupled plates ($d_1\ll a_1$ and $d\gg d_1$, green region in Fig.~\ref{fig:regimes}), the RPA becomes accurate and the capacitance is independent of density, as given by Eq.~(\ref{eq:C0}). The orange region in Fig.~\ref{fig:regimes} depicts a domain where the density is large enough for the electrons to remain delocalized, and the capacitor thick enough for the interplate effects to be weak. This last condition requires $\Ck>\CG$, which is equivalent to $d>a_1$, such that the orange region lies below the diagonal of the diagram. Here we focus on this domain and we model the quantum capacitance to study how $\CQ$ departs from the RPA result upon approaching the limit of localized dipoles.

We consider a regime where cross screening is weak compared to intraplate screening, $|\Pi_{12}|\ll\Pi_{11}$, and the renormalized kinetic capacitance $\Ck^*\equiv e^2\Pi_{11}$ is large compared to the geometrical capacitance. Under these conditions, Eqs.~(\ref{eq:lalphabeta}) and (\ref{eq:W0}) show that $\Pi_{11}$ is determined at leading order by the screening length $\ell_{11}$, according to $\Pi_{11}=2\epsilon/(e^2\ell_{11})$. By solving Eq.~(\ref{eq:lalphabeta}) for $\Pi_{12}$ and substituting the leading-order expression of $\Pi_{11}$, we find that $\Pi_{12}$ is determined by $\ell_{11}$ as well:
	\begin{equation}
		\Pi_{12}\approx\frac{\epsilon}{e^2d}\left(\sqrt{1+\frac{2d}{\ell_{11}}}-1\right).
	\end{equation}
This model of cross-polarizability has the expected properties. $\Pi_{12}$ drops at large $d$, as $\ell_{11}$ approaches $\ell_1$ [Eq.~(\ref{eq:lalpha})], which is independent of $d$. It is furthermore smaller than $\Pi_{11}$, as required by our assumptions. Clearly $\Pi_{12}>0$, which means that cross screening increases the capacitance.

We now use the approximation $\ell_{11}\approx\ell_1$. In doing so, we assume that $\Pi_{11}$ is dominantly a property of the isolated plates, thus neglecting interplate contributions to $\Pi_{11}$ that may be of the same order as $\Pi_{12}$. We expect this approximation to fail when $d\lesssim\ell_1$. The quantum capacitance follows from Eq.~(\ref{eq:C-same}):
	\begin{equation}\label{eq:model1}
		\CQ\approx\frac{\epsilon}{\ell_1}\left[1+\frac{\ell_1}{2d}
		\left(\sqrt{1+\frac{2d}{\ell_1}}-1\right)\right].
	\end{equation}
The condition $\Ck^*>\CG$ becomes $d>\ell_1/2$, which sets the regime of validity of Eq.~(\ref{eq:model1}). Hence, in this regime, positive cross screening effects can increase the quantum capacitance by as much as a factor of $\sqrt{2}$. This $\sim40\%$ enhancement, reached when $d=\ell_1/2$, could easily be detected experimentally.

Having seen that $\Pi_{12}$ is positive and can have a significant effect on the capacitance, we turn to the density dependence, which requires modeling the density dependence of $\ell_1$. In the RPA, the value of $\ell_1$ is $\ell_1^{\mathrm{TF}}=2\epsilon/(e^2\nu_1)$, which for a parabolic band is simply $\ell_1^{\mathrm{TF}}=a_1$. The RPA screening length is reliable only at high density, where exchange and correlation effects disappear \cite{Mahan}. The STLS theory provides a way to address the departure from this limiting value. In this theory, the susceptibility is expressed in terms of a local-field factor $G(q)$, which captures the effect of exchange and correlation on the static screening at different wavelengths (see Appendix~\ref{app:STLS}). The function $G(q)$ is determined by a self-consistency condition that can be solved numerically. Approximate expressions have been proposed, based on the observation that $G(q)$ is related to the pair distribution function. The Hubbard approximation, in particular, uses the exchange-only form of the pair distribution function and reads $1/G(q)=2\sqrt{1+(k_{\mathrm{F}}/q)^2}$, where the factor 2 is for the spin degeneracy \cite{Hubbard-1958, Jonson-1976, Mahan}. Local-field factor and polarizability are two interchangeable ways of representing the susceptibility and they are related by $\Pi(q)=-\chi_0(q)/[1+V(q)G(q)\chi_0(q)]$. By using the Hubbard approximation for $G$, we thus obtain a density-dependent approximation for $\Pi$ and $\ell_1$. We evaluate the screened interaction as $W(0)=\lim_{q\to 0}V(q)/[1+V(q)\Pi(q)]$ after substituting the expression of $\Pi(q)$ in terms of $G(q)$, using the non-interacting susceptibility $\chi_0(0)=-\nu_1$ and $k_{\mathrm{F}}=\sqrt{2\pi n_1}$, and we define the screening length like in Eq.~(\ref{eq:lalphabeta}) as $\ell_1=2\epsilon W(0)/e^2$. The result is
	\begin{equation}\label{eq:l1}
		\ell_1\approx a_1-\frac{d_1}{2\sqrt{2}}=a_1-\frac{1}{\sqrt{8\pi n_1}}.
	\end{equation}
Hence Coulomb interaction reduces $\ell_1$ relative to the RPA value.

\begin{figure}[tb]
\includegraphics[width=0.8\columnwidth]{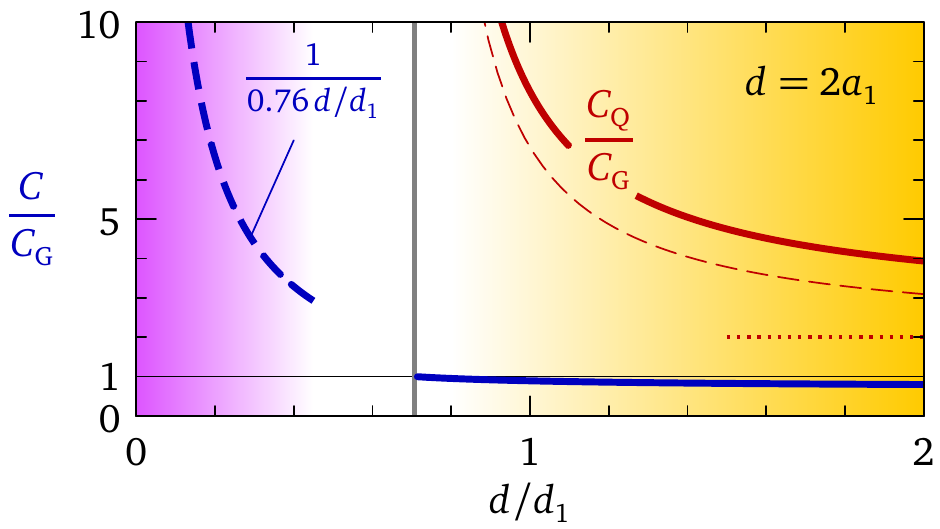}
\caption{\label{fig:identical-plates}
Quantum capacitance (red) and total capacitance (blue) normalized to the geometrical capacitance, versus dimensionless density for a model of symmetric capacitor, Eqs.~(\ref{eq:model1}) and (\ref{eq:l1}) with $d=2a_1$. The dotted line shows the RPA value and the dashed line shows the quantum capacitance without interplate effects. The dashed blue line on the left is the asymptotic result of Ref.~\onlinecite{Skinner-2010}.
}
\end{figure}

In Fig.~\ref{fig:identical-plates}, we plot $\CQ/\CG$ and $C/\CG$ versus $d/d_1$ for a capacitor thickness $d=2a_1$. This value of $d/a_1$ corresponds to the dashed line in Fig.~\ref{fig:regimes} and ensures that the condition $d>\ell_1/2$ is met at all densities. The quantum capacitance (solid red curve) is larger than the RPA value (dotted line) and increases with reducing density, i.e., the system behaves more and more classically until $\ell_1$ vanishes when $d_1=2\sqrt{2}a_1$. The dashed line is the result obtained with $\Pi_{12}=0$, which is lower than the quantum capacitance by a density-dependent factor of $\sim15$--$30$\% for this choice of the thickness. The total capacitance $C$ (blue line) is smaller than $\CG$ and approaches this value as $\CQ$ diverges. For comparison, the figure also shows the total capacitance obtained in Ref.~\onlinecite{Skinner-2010} in the regime of dilute localized dipoles.

We conclude that, in the symmetric capacitor with homogeneous electron densities, the effect of $\Pi_{12}$ is to increase the quantum capacitance and thereby the total capacitance. Note that the assumptions of our model imply $\CQ\gg\CG$, such that $C\approx\CG$ in this regime and the cross quantum capacitance correction, although significant for $\CQ$, is small for $C$. Stronger cross quantum capacitance effects on $C$ may be expected in the ultrathin limit, where $\CQ\ll\CG$ and $C\approx\CQ$. In this limit, however, the interplate correction to $\Pi_{11}$ must be taken into account and modeling is more challenging.

Our model is not smoothly connected to the localized regime, in which $C\gg\CG$ and $\CQ$ is large and negative. As discussed in Sec.~\ref{sec:W}, we expect that the passage from $C<\CG$ to $C>\CG$ is associated with a phase transition, where either $\CQ$ diverges and $C$ is continuous, or $\CQ$ changes sign without diverging and $C$ is discontinuous. In the model, $\CQ$ diverges at the density $n_1^*=1/(8\pi a_1^2)$, where the screening length vanishes within the Hubbard approximation to the STLS theory [Eq.~(\ref{eq:l1})]. The density $n_1^*$ is similar to the density $n_1^{\mathrm{FM}}=1/(2\pi^3a_1^2)$ at which Hartree--Fock theory predicts the two-dimensional electron gas to enter the ferromagnetic state \footnote{The ferromagnetic transition occurs at $r_s=\pi/\sqrt{2}$ \cite{Note2}, with $r_s$ the interparticle distance expressed in units of $a_0=2a_1$.}. Both $n_1^*$ and $n_1^{\mathrm{FM}}$ are much larger than the density at which Wigner crystallization is expected in the isolated plates, i.e., $n_1^{\mathrm{WC}}=1/(4\pi r_s^2a_1^2)$ with $r_s\approx 31$ \cite{Drummond-2009}. If the first instability occurring in the capacitor when reducing density is indeed towards a homogeneous ferromagnetic state, this would imply that at least two distinct phase transitions take place to connect our model to the regime in which a description based on localized electrons is appropriate.

\subsection{Dissimilar plates: Diagrammatic approach}

The model discussed in the previous section leads to a positive $\Pi_{12}$ for identical plates, but does not offer an intuitive picture of the physical mechanism causing this behavior. In order to better understand the origin of $\Pi_{12}$, one has to identify the microscopic processes that contribute to the interplate polarizability. Diagrammatic perturbation theory provides a recipe for calculating the polarizability order by order in the interplate interaction and is therefore a logical route to follow. This approach allows us to relax the assumption of identical plates and to investigate the sign of $\Pi_{12}$ for plates that have different densities and/or different carrier masses. It furthermore allows us to approach the limit of ultrathin capacitors, that is beyond the scope of the previously discussed model. To be analytically tractable, the diagrammatic theory requires approximations, not only in the selection of diagrams, but also in their evaluation. We expect that the results are nevertheless qualitatively correct because---when applied to symmetric capacitors in the appropriate regime---the diagrammatic result predicts orders of magnitude and trends similar to those found with the model of Sec~\ref{sec:symmetric}.

The diagrammatic analysis shows that two interplate polarization mechanisms can be identified. The first mechanism results from Coulomb correlations that amount to an effective repulsion between same-sign charges on opposite plates. These correlations lead to a ``negative'' cross screening that reinforces the intraplate screening and reduces the capacitance as sketched in Fig.~\ref{fig:dQ}. The second mechanism involves Coulomb correlations that produce another---density-dependent---effective interaction, that becomes attractive and leads to ``positive'' cross screening when the equilibrium densities in the plates are equal or nearly equal. The analysis shows that the first process dominates and $\Pi_{12}$ is negative in most of the five-dimensional parameter space defined by the capacitor thickness $d$, the carrier densities $n_1$ and $n_2$, and the carrier masses $m_1$ and $m_2$. $\Pi_{12}$ is positive only in a narrow slice where $n_1$ and $n_2$ are very close to one another, which is the optimal condition for the condensation of all carriers into bound pairs.

\begin{figure}[b]
\includegraphics[width=0.7\columnwidth]{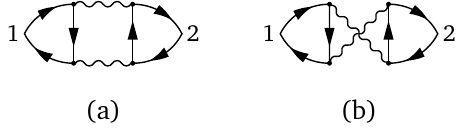}
\caption{\label{fig:diagrams}
Lowest-order interplate polarizability diagrams in (a) the particle-hole and (b) the particle-particle channels.
}
\end{figure}

The details of the diagrammatic calculation are reported in Appendix~\ref{app:diagrams}. Here, we only briefly outline the main steps. There are two diagrams contributing to $\Pi_{12}$ at leading (second) order in the interplate interaction $V_{12}$. They describe polarization processes in the particle-hole [Fig.~\ref{fig:diagrams}(a)] and particle-particle channels [Fig.~\ref{fig:diagrams}(b)], respectively. Similar processes are identified at all subsequent orders, forming two geometric series. An approximate evaluation of these series is performed, where, first, the bare long-range interaction $V_{12}$ is replaced by the short-range screened interaction $W_{12}$, which is furthermore approximated by its value at $q=\omega=0$. The two series of diagrams then lead to emerging effective interactions in the particle-hole and particle-particle channels, that are also replaced by their long-wavelength static values. The resulting expressions can be fully evaluated for parabolic bands at zero temperature and give, up to second order in the screened interaction:
	\begin{multline}\label{eq:Pi12}
		\Pi_{12}\approx\nu_1\nu_2\frac{\mu W_{12}^2}{\pi\hbar^2}\left\{
		-\frac{m_1+m_2}{m_1-m_2}\ln\left(\frac{m_1}{m_2}\right)\right.\\ \left.
		+\ln\left[\frac{1}{\pi d^2}\frac{n_1\left(1+m_2/m_1\right)
		+n_2\left(1+m_1/m_2\right)}{(n_1-n_2)^2}\right]\right\}.
	\end{multline}
$\nu_{\alpha}$ is the DOS in plate $\alpha$, which has equilibrium density $n_{\alpha}$ and mass $m_{\alpha}$, and $\mu=m_1m_2/(m_1+m_2)$ is the reduced mass. The first term is the contribution of the particle-hole channel. This term is negative and varies with varying capacitor thickness via the screened interaction $W_{12}$. The second term is the contribution from the particle-particle channel. It has a peculiar dependence on densities and thickness, which relates to the existence of an interlayer bound state. The singular behavior $\sim\ln(1/d^2)$ results from a simplified treatment of the bound-state energy (see Appendix~\ref{app:diagrams}). This second term can be positive or negative, such that the particle-hole and particle-particle channels compete to set the final sign of $\Pi_{12}$.

To analyze Eq.~(\ref{eq:Pi12}), it is convenient to introduce four dimensionless variables. To this end, we measure all lengths in units of the effective Bohr radius in plate~1, $a_1=2\pi\epsilon\hbar^2/(m_1e^2)$. The parameter $x=d/a_1$ represents the capacitor thickness, while the parameter $y=(8\pi n_1a_1^2)^{1/2}$ represents the density in plate~1 (the factor $8\pi$ is introduced for convenience in order to simplify the formulas.) The difference between the plates is quantified by the parameters $u=m_2/m_1$ and $v=(n_2/n_1)^{1/2}$.

Because $W_{12}$ depends on $\Pi_{12}$ and $\Pi_{21}$, a self-consistent loop must be solved. If we start the loop with $\Pi_{12}=\Pi_{21}=0$, we have $W_{12}=W_{21}$ at the first iteration and the property $\Pi_{12}=\Pi_{21}$ continues to hold until self-consistency. We restrict here to this type of solution. For the diagonal polarizabilities, we use the approximation $\Pi_{\alpha\alpha}\approx 2\epsilon/(e^2\ell_{\alpha\alpha})$ introduced previously with $\ell_{\alpha\alpha}\approx\ell_{\alpha}^{\mathrm{TF}}-1/\sqrt{8\pi n_\alpha}$. Since $\ell_1^{\mathrm{TF}}=a_1$, these approximations take the form $\ell_{11}/a_1\approx1-1/y$ and $\ell_{22}/a_1=1/u-1/vy$. The conditions $\ell_{11}>0$ and $\ell_{22}>0$ require $y>\max(1,u/v)$. In terms of the introduced dimensionless variables, these approximations lead to the following expressions for the polarizabilities:
	\begin{subequations}\label{eq:Pid}\begin{align}
		\frac{\Pi_{11}}{\nu_1}&=\frac{y}{y-1},\qquad
		\frac{\Pi_{22}}{\nu_1}=\frac{uy}{y-u/v}\\
		\nonumber
		\frac{\Pi_{12}}{\nu_1}&=\frac{\Pi_{21}}{\nu_1}=\left(W_{12}\nu_1\right)^2\frac{u^2}{1+u}
		\left\{\frac{u+1}{u-1}\ln\left(\frac{1}{u}\right)\right.\\
		\label{eq:Pidb}
		&\quad\left.+\ln\left[\frac{8(u+1)(u+v^2)}{u(1-v^2)^2(xy)^2}\right]\right\}\\
		\label{eq:Pidc}
		W_{12}\nu_1&=\frac{1+2x\frac{\Pi_{12}}{\nu_1}}{\frac{uy(y-1+y/u-1/v+2xy)}{(y-1)(y-u/v)}
		-2\frac{\Pi_{12}}{\nu_1}-2x\left(\frac{\Pi_{12}}{\nu_1}\right)^2}.
	\end{align}\end{subequations}

It should be noted that---as we have restricted Eq.~(\ref{eq:Pi12}) to the leading order in $W_{12}$---these coupled equations cannot be trusted if $W_{12}$ grows during self-consistency. This only occurs in a narrow parameter range, where $\Pi_{12}$ is large and positive. We show below that $\Pi_{12}$ is negative in most of the parameter space, such that $W_{12}$ decreases during the self-consistent loop. In Appendix~\ref{app:comparison}, we apply these equations to identical plates and show that they provide results for the quantum capacitance that are consistent with those presented in Sec~\ref{sec:symmetric}. This gives us confidence in the simplifications made for evaluating the diagrams and in the ability of Eq.~(\ref{eq:Pid}) to qualitatively describe capacitors with dissimilar plates.

The sign of $\Pi_{12}$ is the sign of the term inside the curly braces in Eq.~(\ref{eq:Pidb}). By regrouping the logarithms, we readily find that the condition for a negative $\Pi_{12}$ is
	\begin{equation}\label{eq:Pi12negative}
		u^{\frac{2u}{1-u}}\frac{(u+1)(u+v^2)}{(1-v^2)^2}
		<{\textstyle\frac{1}{8}}(xy)^2\equiv\left(d/d_1\right)^2,
	\end{equation}
where we have used the interparticle distance $d_{\alpha}=1/\sqrt{\pi n_{\alpha}}$. The right side of the inequality compares the density in plate~1 with the capacitor thickness. We see that the ratio $d/d_1$ is one of the keys that controls the sign of the cross polarizability: $\Pi_{12}>0$ and increase of the quantum capacitance is more likely when this ratio is small, i.e., when the thickness is small compared with interparticle distance---provided that the left side of the inequality is not smaller. The left side of the inequality only depends on the asymmetry of the capacitor regarding mass and density imbalance between the plates \footnote{The asymmetry of Eq.~(\ref{eq:Pi12negative}) regarding the roles of plates~1 and 2 is due to our choice of measuring lengths in units of $a_1$ rather than $a_2$}. This term is large for plates with similar equilibrium densities, $v=d_1/d_2\approx1$, such that the inequality cannot be met, $\Pi_{12}>0$, and interplate effects increase the quantum capacitance relative to the RPA value. This is consistent with the findings of Sec.~\ref{sec:symmetric} (see discussion in Appendix~\ref{app:comparison}).

\begin{figure}[tb]
\includegraphics[width=0.8\columnwidth]{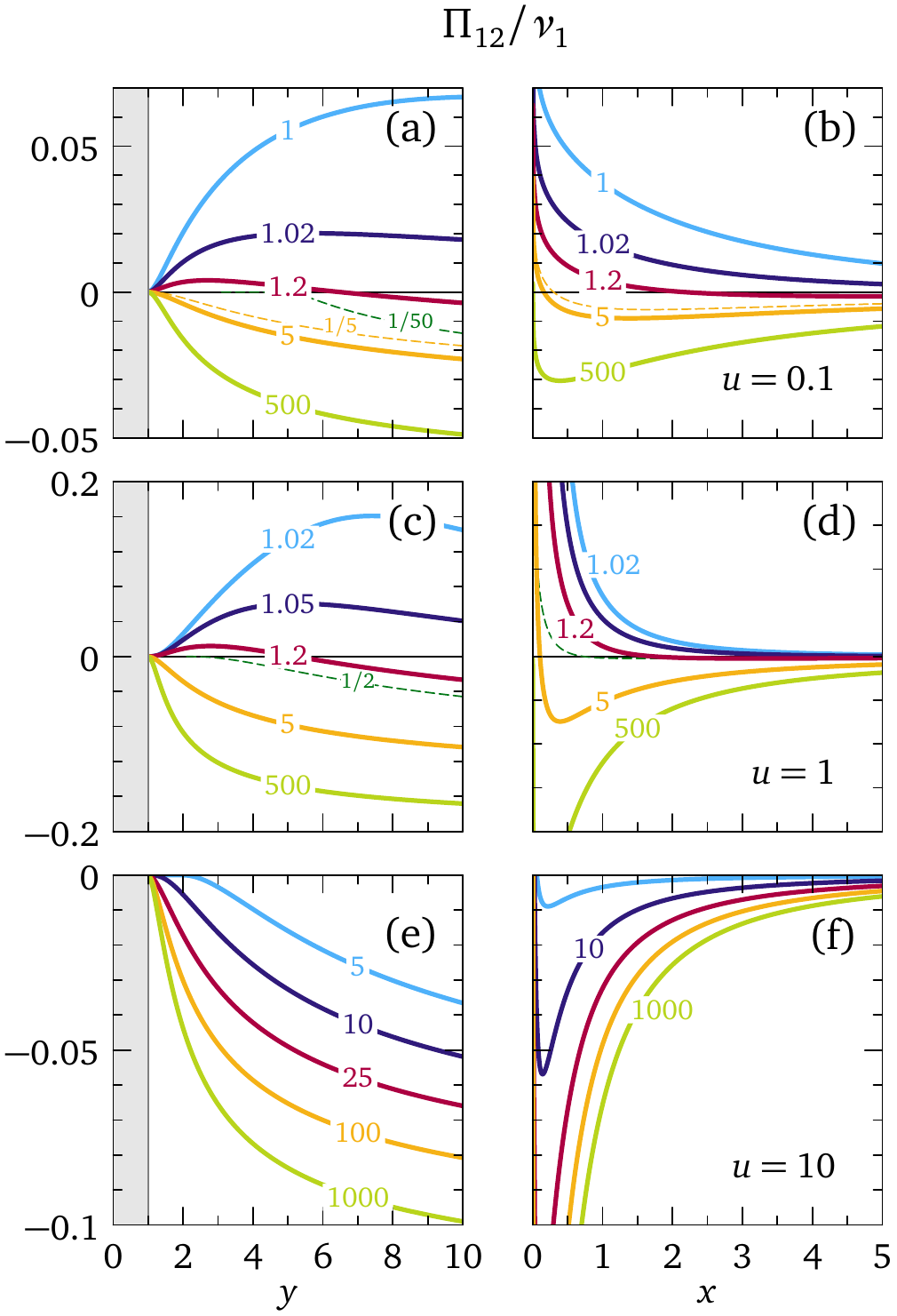}
\caption{\label{fig:Pi12}
Interplate polarizability as given by Eq.~(\ref{eq:Pid}) vs $y$ (dimensionless density of plate~1), $x$ (dimensionless capacitor thickness), $u$ (carriers mass imbalance), and $v$ (carriers density imbalance). $\Pi_{12}$ is plotted vs $y$ at $x=1$ in panels (a), (c), and (e) and versus $x$ at $y=3$ in panels (b), (d), and (f). Panels (a) and (b) correspond to $u=0.1$, (c) and (d) to $u=1$, (e) and (f) to $u=10$. The numbers on the curves show the value of $v$ ($v=1$ means $v=1.001$). The curve for $v=1/50$ in (a) starts at $y=5$, because the physical parameter range is bounded by the condition $y>\max(1,u/v)$.
}
\end{figure}

In order to reach the negative cross screening regime with $\Pi_{12}<0$, we need to maximize the right side---density and/or thickness must be large---and minimize the left side---$v$ must be far from unity. As the amplitude of $\Pi_{12}$ scales like $1/x^2$, in maximizing the right side without loosing interplate effects, it is better to keep $x$ small and rather increase $y$. This is a regime of ultrathin capacitors with not too low densities. For large $y$, the amplitude of $\Pi_{12}$ approaches $1/[(u+1)(2x+1+1/u)^2]$, independent of $v$, such that one can freely choose $v$ away from unity. In summary, the diagrammatic result suggests that the conditions for a negative interplate polarizability are high and different equilibrium densities in the two capacitor plates, and a minimal separation between the plates.

The previous analysis may leave the reader with the impression that $\Pi_{12}<0$ requires fine tuning. This is not the case. We show now that, on the contrary, it is the property $\Pi_{12}>0$ that requires fine tuning. In terms of our dimensionless variables, the five-dimensional parameter space reduces to the four dimensions $(x,y,u,v)$. Figure~\ref{fig:Pi12} shows the behavior of $\Pi_{12}$ in this four-dimensional space. The variation with $y$ (density) is visible in the first column [(a), (c), (e)], the variation with $x$ (thickness) in the second column [(b), (d), (f)], the variation with $u$ (mass imbalance) in the different lines [(a), (b); (c), (d); (e), (f)], and the variation with $v$ (density imbalance) in the different curves in each panel. The main message of the figure is that, as soon as the densities in both plates differ (the numbers on the curves depart from unity), $\Pi_{12}$ turns negative, irrespective of the mass imbalance, density and thickness. The figure also shows that the interplate polarizability is generally larger at higher density and for carriers of equal masses.

\begin{figure}[tb]
\includegraphics[width=0.8\columnwidth]{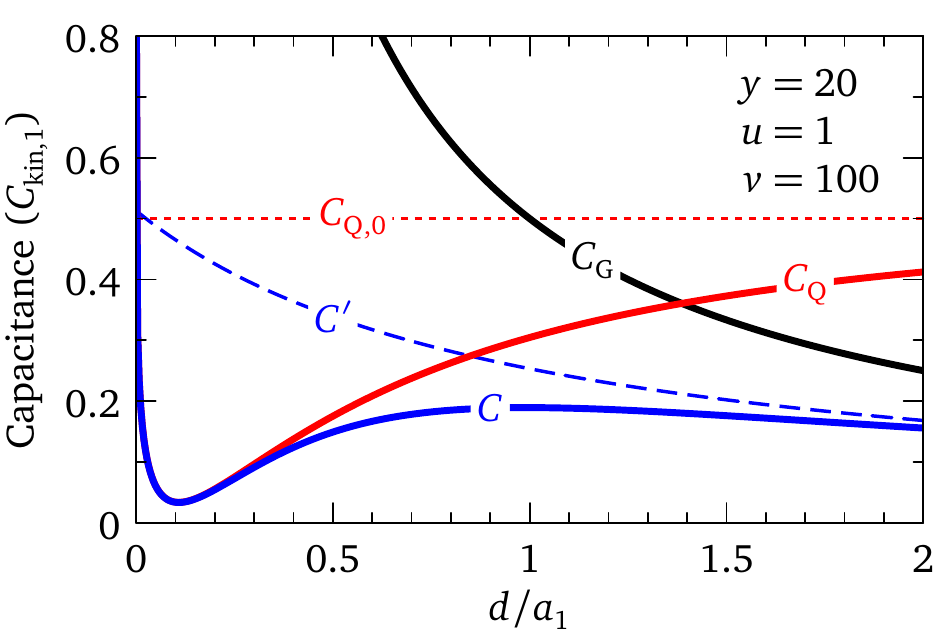}
\caption{\label{fig:nonmonotonic}
Total ($C$, blue), geometrical ($\CG$, black), and quantum ($\CQ$, red) capacitances in units of the kinetic capacitance of plate~1 vs capacitor thickness for $y=20$, $u=1$, and $v=100$. The RPA value of the quantum capacitance ($C_{\mathrm{Q},0}$, dotted-red) and the total capacitance without cross screening ($C'$, dashed-blue) are shown for comparison. $y=(8\pi n_1a_1^2)^{1/2}$ measures the density with $a_1$ the effective Bohr radius, $u=m_2/m_1$ the ratio of carrier masses, and $v=(n_2/n_1)^{1/2}$ the ratio of carrier densities. We used the non-self consistent model in this figure for simplicity. The self-consistent solution shows a similar but slightly weaker effect.
}
\end{figure}

Rather interestingly, the strong suppression of $\Pi_{12}$ with increasing capacitor thickness [Figs.~\ref{fig:Pi12}(b), \ref{fig:Pi12}(d), and \ref{fig:Pi12}(f)] makes it possible to have a situation where the detrimental effect of $\Pi_{12}$ on the total capacitance drops faster with increasing $d$ than the geometrical capacitance, such that the total capacitance actually \emph{increases} with increasing $d$. Figure~\ref{fig:nonmonotonic} shows the evolution of the various capacitances versus thickness in such a case, where the total capacitance has a non-monotonic dependence on thickness. In dimensionless units, the geometrical capacitance is $1/(2x)$ and the noninteracting (RPA) value of the quantum capacitance is $1/(1+1/u)$. The figure also shows the total capacitance $C'$ calculated without cross screening, which gives $1/[2x+1+1/u-(1+1/v)/y]$. Experimentally, a capacitance increasing with increasing $d$---everything else being equal---would be a clear demonstration of cross quantum capacitance, since all other known contributions to the capacitance are either independent of $d$ or decrease with increasing $d$. 

To conclude, we briefly return to Fig.~\ref{fig:dQ}, where an interpretation of the cross quantum capacitance in terms of screening charges is illustrated. The excess charge is the superposition of what would be given by the laws of classical electrostatics and an opposite screening charge. The screening charge is divided into intraplate and interplate terms. The diagrammatic approach underlines that the interplate term has its roots in nonclassical correlations induced by the Coulomb interaction between charges in different plates. A vivid illustration of this is that $\Pi_{12}$ is of second order in the interplate interaction: It follows that the sign of $\Pi_{12}$ does not depend on whether the bare charges attract or repel between the plates. One possibility is that the screening charges in one plate tend to drag opposite-sign charges on the other plate. In Fig.~\ref{fig:dQ}, this is represented as blue charges accumulating in front of opposite orange charges. The net effect is a reinforcement of the screening and a decrease of the total charge. This effective attraction between opposite-sign charges results from the particle-hole polarization diagrams. In the other case---when the particle-particle polarization diagrams win---the screening charges in one plate drag same-sign charges on the other plate, opposing to the intralayer screening, such that the capacitance is enhanced. Similar considerations apply to the physics of Coulomb drag in transport, and it would be interesting to investigate in more details the connection between this phenomenon and the cross quantum capacitance.

\section{Summary and conclusions}

Experimental progress in controlling matter at the atomic scale is now allowing investigations of the electrical capacitance of nanostructures in which two conductors are separated by atomically thin insulators, whose geometrical capacitance is extremely large. Since different contributions to the total capacitance add in series, such a large geometrical capacitance very strongly enhances the experimental sensitivity to all other contributions. Additionally, atomically thin insulators facilitate reaching the regime in which the average distance between charges on one of the capacitor plates is comparable to the distance between charges on opposite plates. In this regime, interaction-induced correlations between charges on opposite plates cannot be neglected.

Virtually all past theoretical studies of the capacitance of nanostructures have discussed specific phenomena, and followed dedicated approaches suitable to their description. The effect of interactions between charges on opposite plates, for instance, has been considered in a series of different papers, but it was analyzed in cases in which it was appropriate to neglect the wave nature of electrons. Other work focusing on quantum capacitance analyzed phenomena originating from the finite density of states and intra-plate interaction effects, but completely disregarded correlations between charges on the two opposite capacitor plates. In the structures that are becoming available for experimental investigations, however, both interplate correlations and the wave nature of electrons need to be considered simultaneously to enable a suitable description.

Our paper develops a general theoretical framework that accounts for the wave nature of electrons, and for both inter and intralayer interaction effects. We have used linear response theory to derive a formula of general validity, in which the total capacitance is expressed as the series connection of the conventional geometrical capacitance and of a second term, which depends exclusively on the intraplate and on the interplate electron polarizabilities. This term describes both the conventional quantum capacitance contribution, which is associated to the intraplate polarizability, and interplate correlations, which lead to a new effect that we refer to as cross quantum capacitance. The formula reduces to the established expression for quantum capacitance in known limits when interplate correlations can be neglected. In capacitors with small electrode separation, however, interplate correlations cannot be neglected, and the cross quantum capacitance can become as important as the conventional quantum capacitance in determining the total capacitance.

As the origin of the cross quantum capacitance is rooted in interaction-induced quantum correlations between charges on opposite plates, the phenomenon has no classical analog and it is difficult to develop an intuition. Even the sign of the effect cannot be easily determined a priori. To gain understanding, we have calculated the interplate polarizability using two models, based on different approximations. Despite being based on very different theoretical approaches, the two models lead to consistent behavior when applied to the same parameter regime, which gives us confidence in the validity of our results. We find that the sign of the interplate polarizability can be either positive or negative. In our calculations, a positive sign---corresponding to an enhancement of the total capacitance relative to the case in which interplate correlations are not included---is found for symmetric electrodes (same electron density and effective mass); a negative sign---which reduces the total capacitance---is common in all other cases. Our calculations further allow us to identify a specific signature of the cross quantum capacitance, namely a non-monotonic dependence of the total capacitance on the separation of the capacitor plates that---if observed experimentally---would provide the ultimate evidence for the relevance of the phenomenon.

Detailed experimental studies of the capacitance of structures in which the effects discussed here can dominate are only starting now, and first work indeed indicates that these effects can be very large. This is the case of ionic liquid gated semiconducting transition metal dichalcogenide monolayers, whose extremely large geometrical capacitance ($\approx 50\mu$F/cm$^2$) makes the contribution related to the finite density of states dominate the total capacitance \cite{Ye-2011, Ji-2014, Zhang-2019}. A quantitative analysis suggests that this contribution cannot be described in terms of the usual quantum capacitance, which motivated us to introduce the notion of cross quantum capacitance analyzed in detail in this paper. Similar considerations can be made for earlier work on ionic gated graphene mono and multilayers \cite{Ye-2011, Kim-2014}.

Clearly, ionic gated devices cannot be directly treated with our theory, which assumes both capacitor plates to be described by a quantum electron liquid, and further theoretical work may be needed. However, the basic concept of cross quantum capacitance that we have developed here remains valid. Based on the magnitude of the effects observed in the experiments \cite{Zhang-2019}, we conclude that the influence of the cross quantum capacitance on the total capacitance can be very large. Indeed, in the configuration of those experiments, it was inferred that the cross quantum capacitance term suppresses the total capacitance by factor of 2-to-3, i.e., the cross quantum capacitance can give a dominating contribution. Other candidate systems for which the theory developed here---or some straightforward variation of it---can be applied and a large cross quantum capacitance effect can be envisioned are conducting 2D materials separated by atomically thin insulators, such as few layer hBN crystals, in which the small thickness of the dielectric results in a very large geometrical capacitance.

To conclude, we note that understanding these phenomena is not only of fundamental interest, but it has great relevance for technology. Indeed, similar effects likely play a role in ionic batteries and limit the total amount of charge that can be stored. In that context, finding how to increase the capacitance by a factor of 2-to-3 would represent a technological breakthrough. That is why developing a much better understanding of the physics determining the total capacitance of structures in which charges on opposite plates are separated by atomic distances is extremely important also for future technological developments.

\begin{acknowledgments}
This research was supported by the Swiss National Science Foundation under Division II. A.F.M.\ acknowledges financial support from the EU Graphene Flagship project.
\end{acknowledgments}

\appendix

\section{Non-local screening in a double-layer system}
\label{app:3D}

In this appendix, we study the full three-dimensional dielectric response of a capacitor with two main goals. The first is to justify the two-dimensional approach used in Sec.~\ref{sec:C-chi} and the second is to prepare the tools for the extension of our results to the more general case with $\epsilon_1\neq\epsilon\neq\epsilon_2$.

\subsection{Dielectric function}

The capacitor is invariant by translation in the $(x,y)$ plane and has axial symmetry in the three-dimensional space. The symmetry of the problem commands to work in the reciprocal space of two-dimensional wave vectors $\vec{q}$ for the in-plane coordinates and in real space for the $z$ direction. We consider an arbitrary test charge distribution $\rho_{\mathrm{ext}}(\vec{q},z)$, to which the capacitor responds with an induced charge $\rho_{\mathrm{ind}}(\vec{q},z)$. The dielectric function is defined as
	\begin{align}\label{eq:rhotot}
		\nonumber
		\rho_{\mathrm{tot}}(\vec{q},z)&=\rho_{\mathrm{ext}}(\vec{q},z)+\rho_{\mathrm{ind}}(\vec{q},z)\\
		&\equiv \int_{-\infty}^{\infty}dz'\,\epsilon^{-1}(\vec{q},z,z')\rho_{\mathrm{ext}}(\vec{q},z').
	\end{align}
In order to derive an expression for $\epsilon^{-1}$, we need to express $\rho_{\mathrm{ind}}$ explicitly in terms of $\rho_{\mathrm{ext}}$. Within linear response, the induced charge is proportional to the potential $V_{\mathrm{ext}}$ of the external charge, the coefficient of proportionality being the charge susceptibility:
	\begin{equation}\label{eq:rhoind}
		\rho_{\mathrm{ind}}(\vec{q},z)
		=\int_{-\infty}^{\infty}dz'\,\chi_{\rho\rho}(\vec{q},z,z')V_{\mathrm{ext}}(\vec{q},z').
	\end{equation}
Because the mobile charges are confined to the capacitor plates, the susceptibility has the structure
	\begin{multline}\label{eq:chirhorho}
		\chi_{\rho\rho}(\vec{q},z,z')=\chi_{11}(\vec{q})\delta(z)\delta(z')
		+\chi_{12}(\vec{q})\delta(z)\delta(z'-d)\\
		+\chi_{21}(\vec{q})\delta(z-d)\delta(z')+\chi_{22}(\vec{q})\delta(z-d)\delta(z'-d),
	\end{multline}
where the plates 1 and 2 have coordinates $z=0$ and $z=d$, respectively. Note that the $\chi_{\alpha\beta}$ in Eq.~(\ref{eq:chirhorho}) are still \emph{charge} susceptibilities at this stage; they will be reinterpreted as number-density correlation functions below, consistently with Eq.~(\ref{eq:chinn}). We now relate $V_{\mathrm{ext}}(\vec{q},z)$ to $\rho_{\mathrm{ext}}(\vec{q},z)$ using the three-dimensional Poisson equation $V_{\mathrm{ext}}(\vec{k})=\rho_{\mathrm{ext}}(\vec{k})/(\epsilon k^2)$. By Fourier transforming along $k_z$, we find
	\begin{align}\label{eq:Poisson}
		\nonumber
		V_{\mathrm{ext}}(\vec{q},z)&=\int_{-\infty}^{\infty}\frac{dk_z}{2\pi}\,e^{ik_zz}
		\frac{\rho_{\mathrm{ext}}(\vec{q},k_z)}{\epsilon(q^2+k_z^2)}\\
		\nonumber
		&=\int_{-\infty}^{\infty}dz'\,\rho_{\mathrm{ext}}(\vec{q},z')
		\int_{-\infty}^{\infty}\frac{dk_z}{2\pi}\,\frac{e^{ik_z(z-z')}}{\epsilon(q^2+k_z^2)}\\
		&=\frac{1}{2\epsilon q}\int_{-\infty}^{\infty}dz'\,\rho_{\mathrm{ext}}(\vec{q},z')
		e^{-q|z-z'|}.
	\end{align}
This relation allows one to write $\rho_{\mathrm{ind}}(\vec{q},z)$ in a form consistent with Eq.~(\ref{eq:rhotot}) and to deduce the dielectric function by identification:
	\begin{multline}\label{eq:epsilon}
		\epsilon^{-1}(\vec{q},z,z')=\delta(z-z')\\+\frac{1}{2\epsilon q}\Big[
		\chi_{11}(\vec{q})\delta(z)e^{-q|z'|}+\chi_{12}(\vec{q})\delta(z)e^{-q|z'-d|}\\
		+\chi_{21}(\vec{q})\delta(z-d)e^{-q|z'|}+\chi_{22}(\vec{q})\delta(z-d)e^{-q|z'-d|}\Big].
	\end{multline}
The non-locality of the dielectric response is here manifested by the fact that the dielectric function depends on $z$ and $z'$ separately, unlike the vacuum permittivity [first term in the right-hand side of Eq.~(\ref{eq:epsilon})], which depends on $z-z'$.

\subsection{Screening of a point charge}
\label{sec:point-charge}

Consider a point charge $Q$ located on the $z$ axis at coordinate $z_Q$. The density and potential for this charge are
	\begin{equation}\label{eq:point-charge}
		\rho(\vec{q},z)=Q\delta(z-z_Q),\quad V(Q,z_Q;\vec{q},z)=Q\frac{e^{-q|z-z_Q|}}{2\epsilon q},
	\end{equation}
as it follows from Eq.~(\ref{eq:Poisson}). For two identical elementary charges in the same plate ($z=z_Q$) and two opposite charges in different plates ($|z-z_Q|=d$), we recover the potential energies $e^2/(2\epsilon q)$ and $-e^2e^{-qd}/(2\epsilon q)$, respectively, entering the calculation of the geometrical capacitance in Eq.~(\ref{eq:Hartree}). It is tempting but incorrect to compute the screened potential $W(\vec{q},z)$ as $\int_{-\infty}^{\infty}dz'\,\epsilon^{-1}(\vec{q},z,z')V(\vec{q},z')$. In the homogeneous three-dimensional case, we indeed have $W(\vec{q},q_z)=V(\vec{q},q_z)/\epsilon(\vec{q},q_z)$, but here, due to inhomogeneity along $z$, the screened potential must be computed from the screened charge using Poisson's equation~(\ref{eq:Poisson}):
	\begin{equation}\label{eq:Poisson1}
		W(\vec{q},z)=\frac{1}{2\epsilon q}\int_{-\infty}^{\infty}dz'\,\rho_{\mathrm{tot}}(\vec{q},z')
		e^{-q|z-z'|}.
	\end{equation}
Equation~(\ref{eq:rhotot}) with $\rho_{\mathrm{ext}}(\vec{q},z)=Q\delta(z-z_Q)$ gives $\rho_{\mathrm{tot}}(\vec{q},z)=Q\epsilon^{-1}(\vec{q},z,z_Q)$. Making use of Eq.~(\ref{eq:epsilon}), we then get
	\begin{multline}\label{eq:W}
		W(Q,z_Q;\vec{q},z)=V(Q,z_Q;\vec{q},z)\\
		+\frac{Q}{(2\epsilon q)^2}\Big[\chi_{11}(\vec{q})e^{-q|z_Q|}e^{-q|z|}
		+\chi_{12}(\vec{q})e^{-q|z_Q-d|}e^{-q|z|}\\
		+\chi_{21}(\vec{q})e^{-q|z_Q|}e^{-q|z-d|}
		+\chi_{22}(\vec{q})e^{-q|z_Q-d|}e^{-q|z-d|}\Big].
	\end{multline}
This expression gives the potential generated at height $z$ by a charge $Q$ located at height $z_Q$ and screened by the capacitor plates characterized by the susceptibilities $\chi_{\alpha\beta}$.

\subsection{Reduction to two dimensions}

The particular choice $Q=-e$ and $z_Q=0$ in Eq.~(\ref{eq:W}) represents an electron in plate~1, such that $-eW(-e,0;\vec{q},0)$ and $+eW(-e,0;\vec{q},d)$ give the potential energies of a negative test charge in plate~1 and a positive test charge in plate~2, respectively, in the presence of a screened electron in plate~1. Conversely, $Q=+e$ and $z_Q=d$ describes the potential generated by a screened hole in plate~2. We now collect these various situations in a $2\times2$ matrix equation involving intraplate and interplate potentials. We consider energies rather than potentials and thus define ($e=|e|$)
	\begin{align*}
		V_{11}&=-eV(-e,0;\vec{q},0), &V_{12}&=-eV(+e,d;\vec{q},0)\\
		V_{21}&=+eV(-e,0;\vec{q},d), &V_{22}&=+eV(+e,d;\vec{q},d).
	\end{align*}
Analogous definitions hold for the screened counterparts $W_{\alpha\beta}$. For convenience, we also reinterpret $\chi_{\alpha\beta}$ as density-density rather than charge-charge correlation functions, i.e., $\chi_{11}\to e^2\chi_{11}$, $\chi_{12}\to -e^2\chi_{12}$, etc. For instance, Eq.~(\ref{eq:W}) gives for $W_{11}$:
	\begin{multline*}
		\underbrace{-eW(-e,0;\vec{q},0)}_{W_{11}}=\underbrace{-eV(-e,0;\vec{q},0)}_{V_{11}}
		+\underbrace{\frac{e^2}{2\epsilon q}}_{V_{11}}\chi_{11}
		\underbrace{\frac{e^2}{2\epsilon q}}_{V_{11}}\\
		-\underbrace{\frac{e^2}{2\epsilon q}}_{V_{11}}\chi_{12}
		\underbrace{\frac{e^2e^{-qd}}{2\epsilon q}}_{-V_{21}}
		-\underbrace{\frac{e^2e^{-qd}}{2\epsilon q}}_{-V_{12}}\chi_{21}
		\underbrace{\frac{e^2}{2\epsilon q}}_{V_{11}}
		+\underbrace{\frac{e^2e^{-qd}}{2\epsilon q}}_{-V_{12}}\chi_{22}
		\underbrace{\frac{e^2e^{-qd}}{2\epsilon q}}_{-V_{21}}.
		\end{multline*}
It is then straightforward to check that the four cases are recast in the matrix equation
	\begin{multline}\label{eq:2x2}
		\begin{pmatrix}W_{11}&W_{12}\\ W_{21}&W_{22}\end{pmatrix}=
		\begin{pmatrix}V_{11}&V_{12}\\ V_{21}&V_{22}\end{pmatrix}\\
		+\begin{pmatrix}V_{11}&V_{12}\\ V_{21}&V_{22}\end{pmatrix}
		\begin{pmatrix}\chi_{11}&\chi_{12}\\ \chi_{21}&\chi_{22}\end{pmatrix}
		\begin{pmatrix}V_{11}&V_{12}\\ V_{21}&V_{22}\end{pmatrix}.
	\end{multline}
It is useful to stress here that, while the bare interaction obeys $V_{12}=V_{21}$, in general $W_{12}\neq W_{21}$. This is somewhat counterintuitive, but obvious from the definition: $W_{12}$ is the potential energy of an \emph{unscreened} electron in plate~1 in the field of a screened hole in plate~2, while $W_{21}$ is the potential energy of a bare hole in plate~2 in the field of a screened electron in plate~1. If the screening properties differ in the two plates, so do $W_{12}$ and $W_{21}$, as Eq.~(\ref{eq:W0}) explicitly shows. We denote $W$, $V$, and $\chi$ the matrices displayed in Eq.~(\ref{eq:2x2}), which has the same formal structure as the usual diagrammatic equation describing screening \cite{Bruus-Flensberg}, $W=V+V\chi V$. Following the standard practice, we separate the reducible and irreducible contributions to the susceptibility. To this end, we introduce the polarizability matrix $\Pi$, which is \emph{defined} by Eq.~(\ref{eq:chi-diag}). We then check that Eq~(\ref{eq:2x2}) is equivalent to Eq.~(\ref{eq:W-diag}). This justifies the two-dimensional $2\times2$ formulation used in Sec.~\ref{sec:C-chi}.

\subsection{Metallic screening and neutrality}

A conductor in contact with an infinite reservoir of mobile charges screens a test charge perfectly over long distances \cite{Mahan}. This means that the displaced charge cancels the test charge and the total charge given by $\int_{-\infty}^{\infty}dz\,\rho_{\mathrm{tot}}(\vec{q}=\vec{0},z)$ vanishes. The dielectric function Eq.~(\ref{eq:epsilon}) enforces this neutrality property. For a test charge at any height, $\rho_{\mathrm{ext}}(\vec{q},z)=Q\delta(z-z_Q)$, the total charge is $Q\int_{-\infty}^{\infty}dz\,\epsilon^{-1}(\vec{q},z,z_Q)$. Performing the $z$ integration in Eq.~(\ref{eq:epsilon}), inserting the $\chi_{\alpha\beta}$ that solve Eq.~(\ref{eq:chi-diag}) and the Coulomb potentials before taking the limit $q\to 0$, we check that the total charge indeed vanishes. By linearity, any distribution of charges is perfectly screened by the capacitor over long distances. We stress that the interplate polarizabilities are not required for a perfect screening: setting $\Pi_{12}=\Pi_{21}=0$ in the above calculation does not break the neutrality.

\section{Nonuniform dielectric background}
\label{app:epsilon}

\newcommand{\kp}{{k_z}\kern-0.38em'}
\newcommand{\kpp}{{k_z}\kern-0.38em''}
\newcommand{\skp}{{k_z}\kern-0.28em'}

We generalize here our results to the case $\epsilon_1\neq\epsilon\neq\epsilon_2$ (see Fig.~\ref{fig:capacitor}). The Coulomb potential in a system composed of three different dielectric media has been worked out in Ref.~\onlinecite{Barrera-1978}. The intra- and interplate potential energies read in this case:
	\begin{subequations}\label{eq:Valphabeta-new}\begin{align}
		V_{11}&=\frac{e^2}{q}\frac{(\epsilon+\epsilon_2)e^{qd}+(\epsilon-\epsilon_2)e^{-qd}}{\Delta}\\
		V_{22}&=\frac{e^2}{q}\frac{(\epsilon+\epsilon_1)e^{qd}+(\epsilon-\epsilon_1)e^{-qd}}{\Delta}\\
		V_{12}&=V_{21}=-\frac{e^2}{q}\frac{2\epsilon}{\Delta}\\
		\Delta&=(\epsilon_1+\epsilon)(\epsilon+\epsilon_2)e^{qd}
		+(\epsilon_1-\epsilon)(\epsilon-\epsilon_2)e^{-qd}.
	\end{align}\end{subequations}
The Hartree energy in Eq.~(\ref{eq:Hartree}) is $\lim_{q\to0}n^2(V_{11}/2+V_{22}/2+V_{12})$ and remains unchanged, hence the geometrical capacitance Eq.~(\ref{eq:CG}) remains the same as well. We also find that the expressions Eq.~(\ref{eq:W0}) for the screened interaction and Eq.~(\ref{eq:C-Pi}) for the capacitance remain unchanged, as may have been anticipated, since the quantum capacitance in Eq.~(\ref{eq:C-Pi}) does not depend explicitly on $\epsilon$.

We now generalize the analysis of Appendix~\ref{app:3D}. For a nonuniform dielectric constant $\epsilon(z)$, the displacement field is $\vec{D}=\epsilon(z)(-\vec{\nabla}V)$ and the Gauss law $\vec{\nabla}\cdot\vec{D}=\rho$ yields the Poisson equation $-\vec{\nabla}\cdot[\epsilon(z)\vec{\nabla}V]=\rho$. In reciprocal space, this reads
	\begin{equation}
		\int_{-\infty}^{\infty}\frac{d\kp}{2\pi}\,\epsilon(k_z-\kp)(q^2+k_z\kp)V(\vec{q},\kp)=\rho(\vec{q},k_z).
	\end{equation}
Introducing the inverse kernel $\mathscr{I}_q(k_z,\kp)$ defined by
	\begin{equation}\label{eq:kernel-definition}
		\int_{-\infty}^{\infty}\frac{d\kp}{2\pi}\,\epsilon(k_z-\kp)(q^2+k_z\kp)
		\mathscr{I}_q(\kp,\kpp)=2\pi\delta(k_z-\kpp),
	\end{equation}
we relate the potential $V$ to the charge $\rho$ through
	\begin{equation}\label{eq:Poisson-new}
		V(\vec{q},k_z)=\int_{-\infty}^{\infty}\frac{d\kp}{2\pi}\,\mathscr{I}_q(k_z,\kp)\rho(\vec{q},\kp).
	\end{equation}
Equation~(\ref{eq:Poisson-new}) replaces Eq.~(\ref{eq:Poisson}). For a point-charge $Q$ at height $z_Q$, $\rho(\vec{q},k_z)=Qe^{-ik_zz_Q}$ and Eq.~(\ref{eq:Poisson-new}) transforms into
	\begin{equation}\label{eq:point-charge-new}
		V(Q,z_Q;\vec{q},z)=Q\mathscr{I}_q(z,-z_Q),
	\end{equation}
where the kernel $\mathscr{I}_q$ was transformed back to real space. This replaces Eq.~(\ref{eq:point-charge}) and provides a closed expression for the kernel in real space, using the results of Ref.~\onlinecite{Barrera-1978}. The induced charge given by Eq.~(\ref{eq:rhoind}) is replaced by
	\begin{multline}
		\rho_{\mathrm{ind}}(\vec{q},z)=\int_{-\infty}^{\infty}dz'\,\left[\int_{-\infty}^{\infty}dz''\,
		\mathscr{I}_q(z'',-z')\chi_{\rho\rho}(\vec{q},z,z'')\right]\\
		\times\rho_{\mathrm{ext}}(\vec{q},z').
	\end{multline}
Comparing with Eq.~(\ref{eq:rhotot}), we see that the quantity in square brackets yields the response term of the dielectric function. Using Eq.~(\ref{eq:chirhorho}), we thus find that the dielectric function Eq.~(\ref{eq:epsilon}) is replaced by
	\begin{multline}\label{eq:epsilon-new}
		\epsilon^{-1}(\vec{q},z,z')=\delta(z-z')+\left[\chi_{11}(\vec{q})\delta(z)\mathscr{I}_q(0,-z')\right.\\
		+\chi_{12}(\vec{q})\delta(z)\mathscr{I}_q(d,-z')+\chi_{21}(\vec{q})\delta(z-d)\mathscr{I}_q(0,-z')\\
		\left.+\chi_{22}(\vec{q})\delta(z-d)\mathscr{I}_q(d,-z')\right].
	\end{multline}
For a uniform medium $\epsilon(z)\equiv\epsilon$, the solution of Eq.~(\ref{eq:kernel-definition}) or Eq.~(\ref{eq:point-charge-new}) is $\mathscr{I}_q(z,-z')=e^{-q|z-z'|}/(2\epsilon q)$ and Eq.~(\ref{eq:epsilon-new}) recovers Eq.~(\ref{eq:epsilon}). Equation~(\ref{eq:epsilon-new}) warrants the neutrality condition $\int_{-\infty}^{\infty}dz\,\epsilon^{-1}(\vec{0},z,z_Q)=0$ for any value of $z_Q$.

\section{Classical and quantum charge responses}
\label{app:chiclass}

The notion of polarizability is pivotal for the description of the quantum capacitance proposed in this paper. However, this notion may occasionally hurt the intuition, because it has no classical analog. More precisely, the polarizability is a quantity that becomes infinite rather than zero when the laws of quantum mechanics are taken to their classical limit. Our goal in this appendix is to elucidate this point by proposing an interpretation of the polarizability as a way to separate, in the charge response, the part resulting from classical electrostatics and the part resulting from quantum mechanics. We first derive the susceptibility of the capacitor ignoring all quantum effects. This ``classical susceptibility'' explains the geometrical capacitance and allows one to understand why there is no ``classical polarizability''. We then x-ray Eq.~(\ref{eq:chi-diag}) to distinguish classical and quantum contributions in the total charge response and we show that the latter is wholly accounted for by the polarizability.

The linear-response approach used in Sec.~\ref{sec:C-chi} for a quantum capacitor works equally well for a classical capacitor. The classical linear response to an external potential is given by the Poisson equation. In a three-dimensional system with uniform dielectric constant, this equation is $\rho=-en=\epsilon q^2U$. The classical susceptibility defined through $n=\chi^{\mathrm{class}}U$ is simply $\chi^{\mathrm{class}}=-\epsilon q^2/e$. It vanishes in the long-wavelength limit, because classical charges are immune to a uniform electrostatic potential. In the geometry of a capacitor, the Poisson equation has the form given in Eq.~(\ref{eq:Poisson}). To conform with the sign convention of Sec.~\ref{sec:C-chi}, we define the potential energies on the plates as $U_1(\vec{q})=|e|V_{\mathrm{ext}}(\vec{q},0)$ and $U_2(\vec{q})=-|e|V_{\mathrm{ext}}(\vec{q},d)$, and we write $\rho_{\mathrm{ext}}(\vec{q},z)=-|e|n_1(\vec{q})\delta(z)+|e|n_2(\vec{q})\delta(z-d)$, where here (and only here) $n_1$ and $n_2$ denote the \emph{excess} densities on the plates, that are in general different if $U_2\neq-U_1$. Equation~(\ref{eq:Poisson}) can then be recast in the $2\times2$ matrix form
	\begin{equation}
		\begin{pmatrix}n_1\\[0.2em]n_2\end{pmatrix}=
		\begin{pmatrix}\chi^{\mathrm{class}}_{11}&\chi^{\mathrm{class}}_{12}\\[0.2em]
		\chi^{\mathrm{class}}_{21}&\chi^{\mathrm{class}}_{22}\end{pmatrix}
		\begin{pmatrix}U_1\\[0.2em]U_2\end{pmatrix}
	\end{equation}
with
	\begin{subequations}\label{eq:chi-class}\begin{align}
		\chi^{\mathrm{class}}_{11}=\chi^{\mathrm{class}}_{22}&=-\frac{\epsilon}{e^2}q[1+\coth(qd)]\\
		\chi^{\mathrm{class}}_{12}=\chi^{\mathrm{class}}_{21}&=-\frac{\epsilon}{e^2}\frac{q}{\sinh(qd)}.
	\end{align}\end{subequations}
The intraplate response is finite at $q=0$, increases monotonically with increasing $q$, and becomes linear at large $q$. This contrasts with the three-dimensional case, where the classical susceptibility is exactly proportional to $q^2$. Due to the cutoff $e^{-qd}$ in the interplate Coulomb potential, there is no interplate response at wavelengths much shorter than $d$, such that $\chi^{\mathrm{class}}_{12}$ is a decreasing function that vanishes exponentially at large $q$. Despite these qualitative differences, both intra- and interplate responses approach the same value $-\epsilon/(e^2d)$ in the long-wavelength limit. Inserted in Eq.~(\ref{eq:C1}), the macroscopic classical susceptibilities correctly give $C=\CG$.

One could have derived Eq.~(\ref{eq:chi-class}) directly from the relation $\chi^{\mathrm{class}}=-V^{-1}$, where $V$ is the Coulomb interaction. This relation is obvious in the uniform three-dimensional continuum, where $V=e/(\epsilon q^2)$ and $\chi^{\mathrm{class}}=-\epsilon q^2/e$. Using the $2\times2$ Coulomb-interaction matrix of the capacitor, one checks that this relation holds there as well. In an attempt to define a ``classical polarizability'', one could invert Eq.~(\ref{eq:chi-diag}), which gives $\Pi=-\chi(1+V\chi)^{-1}$. The substitution of $\chi^{\mathrm{class}}=-V^{-1}$ would then yield a $\Pi^{\mathrm{class}}$ that is formally infinite. Physically, infinite polarizability signifies that nothing stops classical charges from moving in an external potential, unlike quantum charges that are limited by the Pauli exclusion. The polarizability of free \emph{quantum} charges is the Fermi-level DOS $\propto m/\hbar^2$, which indeed diverges in the classical limit $\hbar\to0$.

Coming back to the capacitor, the susceptibility $\chi_{\alpha\beta}$ determines how the charges in plate $\alpha$ move in response to an external potential applied in plate $\beta$. This motion is a mixture of classical electrostatics and quantum correlations. The classical motion ultimately leads to the geometrical capacitance and the quantum correlations to the quantum capacitance. The separation between classical and quantum motions is not explicit in Eqs.~(\ref{eq:C1}) and (\ref{eq:C2}), because both effects are mixed in $\chi_{\alpha\beta}$. It is therefore necessary to distinguish these contributions: this is where the polarizability comes into play. The response to the external potential must be self-consistent. The primary displaced charge---by this we mean: the charge that would be displaced, were the Coulomb interaction inexistent---induces via the Coulomb interaction a potential that opposes the external potential and readjusts the response. This is the phenomenon of screening. The primary displaced charge is written $-\Pi_{\alpha\beta}U$, where $\Pi_{\alpha\beta}$ is the polarizability and the minus sign reflects that the primary charge is opposite to the source charge of the external potential $U$. Free quantum charges move in an external potential insofar as empty states are available. The calculation shows that their polarizability $\Pi^0$ is proportional to the Fermi-level DOS (Appendix~\ref{app:diagrams}). Note that this result indeed does not require a Coulomb interaction. The Coulomb interaction has two distinct effects. On the one hand, the uncorrelated response $\Pi^0$ is replaced by a correlated response $\Pi$. The physical intuition is that charges have a reduced probability to be at a short distance from one another (exchange-correlation hole). When one charge moves in response to the external potential, other charges are prevented from moving nearby. It can be seen in the main text that $\Pi^0$ yields the kinetic capacitance, the difference between $\Pi$ and $\Pi^0$ accounting for the quantum corrections. On the other hand, the Coulomb interaction triggers the self-consistent screening, which obeys the Poisson equation and may therefore be labeled ``classical''. Specifically, the primary charge $-\Pi_{\alpha\beta}U$ generates a potential $V_{\alpha\alpha}(-\Pi_{\alpha\beta}U)$ in plate $\alpha$ and $V_{\beta\alpha}(-\Pi_{\alpha\beta}U)$ in plate $\beta$, where $V_{\alpha\alpha}$ and $V_{\beta\alpha}$ are the intra- and interplate Coulomb potentials, respectively (all quantities depend on $\vec{q}$). Note how quantum and classical effects get mixed here: the charge $-\Pi U$ is limited by the Pauli principle and other quantum correlations, but the potential $V(-\Pi U)$ is set by the classical Poisson equation. The system responds to the secondary potential like it does to the external potential, which triggers a geometric series of terms that eventually converges to give the total response $\chi_{\alpha\beta}U$. Hence $\chi_{\alpha\beta}$ rests on a quantum ``kernel'', $\Pi$, that determines the responses to the classical potentials generated at the successive orders. In the diagrammatic jargon, the kernel is called the ``irreducible'' part of the response, which has a simple graphical meaning for the diagrams (see Appendix~\ref{app:diagrams}).

\section{Diagrammatic theory of the cross-polarizability}
\label{app:diagrams}

In this appendix, we sketch a diagrammatic theory of the cross quantum capacitance and we evaluate approximately two important series of diagrams. Our purpose is to show that these series lead to an interplate polarizability proportional to the DOS of the capacitor plates and whose sign can be positive or negative depending on $d$, the masses, and the equilibrium densities of carriers in the plates.

\begin{fmffile}{diagrams/chi}

The static susceptibilities defined in Eq.~(\ref{eq:chinn}) are represented by the following diagram:
	\begin{align}
		\chi_{\alpha\beta}(\vec{q})&=-\hspace{2mm}
		\raisebox{-2mm}{
		\begin{fmfgraph*}(16,6)\fmfstyle
			\fmfleft{o}\fmfright{i}
			\fmfpoly{shaded}{p1,p2,p3,p4}
			\fmffixedy{h}{p1,p2}
			\fmf{phantom}{i,p1}\fmf{phantom}{p4,o}\fmf{phantom}{o,p3}\fmf{phantom}{p2,i}
			\fmffreeze
			\fmfi{fermion}{vloc(__o){1,2} .. tension 1 .. {right}vloc(__p3)}
			\fmfi{fermion}{vloc(__p2){right} .. tension 1 .. {1,-2}vloc(__i)}
			\fmfi{fermion}{vloc(__i){-1,-2} .. tension 1 .. {left}vloc(__p1)}
			\fmfi{fermion}{vloc(__p4){left} .. tension 1 .. {-1,2}vloc(__o)}
			\fmfdot{p1,p2,p3,p4}
			\fmfv{label=${\scriptstyle \alpha}$,label.dist=0.5mm}{o}
			\fmfv{label=${\scriptstyle \beta}$,label.dist=0.5mm}{i}
		\end{fmfgraph*}
		}
	\end{align}
The minus sign cancels the sign of the loop and ensures that the relation is consistent with the diagrammatic rules. The left part of the diagram represents the density operator in plate $\alpha$,
	\begin{equation}
		\raisebox{-2mm}{
		\begin{fmfgraph*}(5,6)\fmfstyle
			\fmfleft{o}
			\fmffixedy{h}{p4,p3}\fmffixedx{w}{o,p3}\fmffixedx{w}{o,p4}
			\fmf{phantom}{p4,o}\fmf{phantom}{o,p3}
			\fmffreeze
			\fmfi{fermion}{vloc(__o){1,2} .. tension 1 .. {right}vloc(__p3)}
			\fmfi{fermion}{vloc(__p4){left} .. tension 1 .. {-1,2}vloc(__o)}
			\fmfv{label=${\scriptstyle \alpha}$,label.dist=0.5mm}{o}
		\end{fmfgraph*}
		}
		\hspace{1mm}=
		n_{\alpha}(\vec{q})=\sum_{\vec{k}\sigma}c^{\dagger}_{\alpha\vec{k}\sigma}
		c^{\phantom{\dagger}}_{\alpha\vec{k}+\vec{q}\sigma},
	\end{equation}
the right part is the density operator in plate $\beta$, and the shaded box represents all ways of connecting the fermion lines by intra- and interplate Coulomb interactions. Equation~(\ref{eq:chi-diag}) corresponds to the diagrammatic equation
	\begin{align*}
		\raisebox{-2mm}{
		\begin{fmfgraph*}(16,6)\fmfstyle
			\fmfleft{o}\fmfright{i}
			\fmfpoly{shaded}{p1,p2,p3,p4}
			\fmffixedy{h}{p1,p2}
			\fmf{phantom}{i,p1}\fmf{phantom}{p4,o}\fmf{phantom}{o,p3}\fmf{phantom}{p2,i}
			\fmffreeze
			\fmfi{fermion}{vloc(__o){1,2} .. tension 1 .. {right}vloc(__p3)}
			\fmfi{fermion}{vloc(__p2){right} .. tension 1 .. {1,-2}vloc(__i)}
			\fmfi{fermion}{vloc(__i){-1,-2} .. tension 1 .. {left}vloc(__p1)}
			\fmfi{fermion}{vloc(__p4){left} .. tension 1 .. {-1,2}vloc(__o)}
			\fmfdot{p1,p2,p3,p4}
		\end{fmfgraph*}
		}
		&=
		\raisebox{-2mm}{
		\begin{fmfgraph*}(16,6)\fmfstyle
			\fmfleft{o}\fmfright{i}
			\fmfpoly{full}{p1,p2,p3,p4}
			\fmffixedy{h}{p1,p2}
			\fmf{phantom}{i,p1}\fmf{phantom}{p4,o}\fmf{phantom}{o,p3}\fmf{phantom}{p2,i}
			\fmffreeze
			\fmfi{fermion}{vloc(__o){1,2} .. tension 1 .. {right}vloc(__p3)}
			\fmfi{fermion}{vloc(__p2){right} .. tension 1 .. {1,-2}vloc(__i)}
			\fmfi{fermion}{vloc(__i){-1,-2} .. tension 1 .. {left}vloc(__p1)}
			\fmfi{fermion}{vloc(__p4){left} .. tension 1 .. {-1,2}vloc(__o)}
			\fmfdot{p1,p2,p3,p4}
		\end{fmfgraph*}
		}
		+
		\raisebox{-2mm}{
		\begin{fmfgraph*}(38,6)\fmfstyle
			\fmfleft{o}\fmfright{i}
			\fmfpoly{full}{p11,p12,p13,p14}
			\fmfpoly{shaded}{p21,p22,p23,p24}
			\fmf{wiggly}{v1,v2}
			\fmffixedy{h}{p11,p12}\fmffixedy{h}{p21,p22}\fmffixedx{16}{v1,v2}
			\fmf{phantom}{i,p21}\fmf{phantom}{p24,v2}\fmf{phantom}{v2,p23}\fmf{phantom}{p22,i}
			\fmf{phantom}{v1,p11}\fmf{phantom}{p14,o}\fmf{phantom}{o,p13}\fmf{phantom}{p12,v1}
			\fmffreeze
			\fmfi{fermion}{vloc(__o){1,2} .. tension 1 .. {right}vloc(__p13)}
			\fmfi{fermion}{vloc(__p12){right} .. tension 1 .. {1,-2}vloc(__v1)}
			\fmfi{fermion}{vloc(__v1){-1,-2} .. tension 1 .. {left}vloc(__p11)}
			\fmfi{fermion}{vloc(__p14){left} .. tension 1 .. {-1,2}vloc(__o)}
			\fmfi{fermion}{vloc(__v2){1,2} .. tension 1 .. {right}vloc(__p23)}
			\fmfi{fermion}{vloc(__p22){right} .. tension 1 .. {1,-2}vloc(__i)}
			\fmfi{fermion}{vloc(__i){-1,-2} .. tension 1 .. {left}vloc(__p21)}
			\fmfi{fermion}{vloc(__p24){left} .. tension 1 .. {-1,2}vloc(__v2)}
			\fmfdot{p11,p12,p13,p14,p21,p22,p23,p24,v1,v2}
		\end{fmfgraph*}
		}
		\\
		-\chi\hspace{2em} &=\hspace{2.3em}\Pi\hspace{2.25em}+\hspace{2.35em}
		\Pi\hspace{0.25em}\times\hspace{0.25em}(-V)\hspace{0.05em}\times\hspace{0.05em}(-\chi)
	\end{align*}
The black box represents only a subset of all connections contained in the shaded box. Indeed, the polarizability $\Pi$ is defined as the sum of diagrams that are irreducible, i.e., that cannot be split in two disconnected pieces by cutting just one interaction line. All reducible diagrams, on the other hand, have the structure shown in the last term of the equation. Equation~(\ref{eq:W-diag}) corresponds to the diagrammatic result
	\begin{align}\label{eq:screening-diagrams}
		\raisebox{-2mm}{
		\begin{fmfgraph*}(8,6)\fmfstyle
			\fmfleft{o}\fmfright{i}
			\fmf{dbl_wiggly}{o,i}
		\end{fmfgraph*}
		}
		&=
		\raisebox{-2mm}{
		\begin{fmfgraph*}(8,6)\fmfstyle
			\fmfleft{o}\fmfright{i}
			\fmf{wiggly}{o,i}
		\end{fmfgraph*}
		}
		+
		\raisebox{-2mm}{
		\begin{fmfgraph*}(32,6)\fmfstyle
			\fmfleft{o}\fmfright{i}
			\fmfpoly{full}{p1,p2,p3,p4}
			\fmf{wiggly}{o,v1}\fmf{dbl_wiggly}{v2,i}
			\fmffixedy{h}{p1,p2}\fmffixedx{48}{v1,v2}
			\fmf{phantom}{v2,p1}\fmf{phantom}{p4,v1}\fmf{phantom}{v1,p3}\fmf{phantom}{p2,v2}
			\fmffreeze
			\fmfi{fermion}{vloc(__v1){1,2} .. tension 1 .. {right}vloc(__p3)}
			\fmfi{fermion}{vloc(__p2){right} .. tension 1 .. {1,-2}vloc(__v2)}
			\fmfi{fermion}{vloc(__v2){-1,-2} .. tension 1 .. {left}vloc(__p1)}
			\fmfi{fermion}{vloc(__p4){left} .. tension 1 .. {-1,2}vloc(__v1)}
			\fmfdot{v1,v2,p1,p2,p3,p4}
		\end{fmfgraph*}
		}
		\\
		\nonumber
		-W\hspace{0.7em} &=\hspace{0.6em}-V\hspace{0.85em}+\hspace{0.2em}
		(-V)\hspace{0.55em}\times\hspace{0.55em}\Pi\hspace{0.45em}\times\hspace{0.45em}(-W)
	\end{align}
The various quantities in the previous diagrammatic relations are $2\times2$ matrices and matrix products are implied. For instance, the explicit expression for $-\chi_{12}$ is
	\begin{align*}
		\raisebox{-2mm}{
		\begin{fmfgraph*}(14.4,5.4)\fmfstyle
			\fmfleft{o}\fmfright{i}
			\fmfpoly{shaded}{p1,p2,p3,p4}
			\fmffixedy{h}{p1,p2}
			\fmf{phantom}{i,p1}\fmf{phantom}{p4,o}\fmf{phantom}{o,p3}\fmf{phantom}{p2,i}
			\fmffreeze
			\fmfi{fermion}{vloc(__o){1,2} .. tension 1 .. {right}vloc(__p3)}
			\fmfi{fermion}{vloc(__p2){right} .. tension 1 .. {1,-2}vloc(__i)}
			\fmfi{fermion}{vloc(__i){-1,-2} .. tension 1 .. {left}vloc(__p1)}
			\fmfi{fermion}{vloc(__p4){left} .. tension 1 .. {-1,2}vloc(__o)}
			\fmfdot{p1,p2,p3,p4}
			\fmfv{label=${\scriptstyle 1}$,label.dist=0.5mm}{o}
			\fmfv{label=${\scriptstyle 2}$,label.dist=0.5mm}{i}
		\end{fmfgraph*}
		}
		\hspace{2mm}=\hspace{2mm}
		\raisebox{-2mm}{
		\begin{fmfgraph*}(14.4,5.4)\fmfstyle
			\fmfleft{o}\fmfright{i}
			\fmfpoly{full}{p1,p2,p3,p4}
			\fmffixedy{h}{p1,p2}
			\fmf{phantom}{i,p1}\fmf{phantom}{p4,o}\fmf{phantom}{o,p3}\fmf{phantom}{p2,i}
			\fmffreeze
			\fmfi{fermion}{vloc(__o){1,2} .. tension 1 .. {right}vloc(__p3)}
			\fmfi{fermion}{vloc(__p2){right} .. tension 1 .. {1,-2}vloc(__i)}
			\fmfi{fermion}{vloc(__i){-1,-2} .. tension 1 .. {left}vloc(__p1)}
			\fmfi{fermion}{vloc(__p4){left} .. tension 1 .. {-1,2}vloc(__o)}
			\fmfdot{p1,p2,p3,p4}
			\fmfv{label=${\scriptstyle 1}$,label.dist=0.5mm}{o}
			\fmfv{label=${\scriptstyle 2}$,label.dist=0.5mm}{i}
		\end{fmfgraph*}
		}
		&\hspace{2mm}+\hspace{2mm}
		\raisebox{-2mm}{
		\begin{fmfgraph*}(34.2,5.4)\fmfstyle
			\fmfleft{o}\fmfright{i}
			\fmfpoly{full}{p11,p12,p13,p14}
			\fmfpoly{shaded}{p21,p22,p23,p24}
			\fmf{wiggly}{v1,v2}
			\fmffixedy{h}{p11,p12}\fmffixedy{h}{p21,p22}\fmffixedx{16}{v1,v2}
			\fmf{phantom}{i,p21}\fmf{phantom}{p24,v2}\fmf{phantom}{v2,p23}\fmf{phantom}{p22,i}
			\fmf{phantom}{v1,p11}\fmf{phantom}{p14,o}\fmf{phantom}{o,p13}\fmf{phantom}{p12,v1}
			\fmffreeze
			\fmfi{fermion}{vloc(__o){1,2} .. tension 1 .. {right}vloc(__p13)}
			\fmfi{fermion}{vloc(__p12){right} .. tension 1 .. {1,-2}vloc(__v1)}
			\fmfi{fermion}{vloc(__v1){-1,-2} .. tension 1 .. {left}vloc(__p11)}
			\fmfi{fermion}{vloc(__p14){left} .. tension 1 .. {-1,2}vloc(__o)}
			\fmfi{fermion}{vloc(__v2){1,2} .. tension 1 .. {right}vloc(__p23)}
			\fmfi{fermion}{vloc(__p22){right} .. tension 1 .. {1,-2}vloc(__i)}
			\fmfi{fermion}{vloc(__i){-1,-2} .. tension 1 .. {left}vloc(__p21)}
			\fmfi{fermion}{vloc(__p24){left} .. tension 1 .. {-1,2}vloc(__v2)}
			\fmfdot{p11,p12,p13,p14,p21,p22,p23,p24}
			\fmfv{label=${\scriptstyle 1}$,label.dist=0.45mm}{o}
			\fmfv{label=${\scriptstyle 2}$,label.dist=0.45mm}{i}
			\fmfv{label=${\scriptstyle 1}$,label.dist=1mm,label.angle=70}{v1}
			\fmfv{label=${\scriptstyle 1}$,label.dist=1mm,label.angle=110}{v2}
			\fmfdot{v1,v2}
		\end{fmfgraph*}
		}
		\\[2mm]
		&\hspace{2mm}+\hspace{2mm}
		\raisebox{-2mm}{
		\begin{fmfgraph*}(34.2,5.4)\fmfstyle
			\fmfleft{o}\fmfright{i}
			\fmfpoly{full}{p11,p12,p13,p14}
			\fmfpoly{shaded}{p21,p22,p23,p24}
			\fmf{wiggly}{v1,v2}
			\fmffixedy{h}{p11,p12}\fmffixedy{h}{p21,p22}\fmffixedx{16}{v1,v2}
			\fmf{phantom}{i,p21}\fmf{phantom}{p24,v2}\fmf{phantom}{v2,p23}\fmf{phantom}{p22,i}
			\fmf{phantom}{v1,p11}\fmf{phantom}{p14,o}\fmf{phantom}{o,p13}\fmf{phantom}{p12,v1}
			\fmffreeze
			\fmfi{fermion}{vloc(__o){1,2} .. tension 1 .. {right}vloc(__p13)}
			\fmfi{fermion}{vloc(__p12){right} .. tension 1 .. {1,-2}vloc(__v1)}
			\fmfi{fermion}{vloc(__v1){-1,-2} .. tension 1 .. {left}vloc(__p11)}
			\fmfi{fermion}{vloc(__p14){left} .. tension 1 .. {-1,2}vloc(__o)}
			\fmfi{fermion}{vloc(__v2){1,2} .. tension 1 .. {right}vloc(__p23)}
			\fmfi{fermion}{vloc(__p22){right} .. tension 1 .. {1,-2}vloc(__i)}
			\fmfi{fermion}{vloc(__i){-1,-2} .. tension 1 .. {left}vloc(__p21)}
			\fmfi{fermion}{vloc(__p24){left} .. tension 1 .. {-1,2}vloc(__v2)}
			\fmfdot{p11,p12,p13,p14,p21,p22,p23,p24}
			\fmfv{label=${\scriptstyle 1}$,label.dist=0.45mm}{o}
			\fmfv{label=${\scriptstyle 2}$,label.dist=0.45mm}{i}
			\fmfv{label=${\scriptstyle 1}$,label.dist=1mm,label.angle=70}{v1}
			\fmfv{label=${\scriptstyle 2}$,label.dist=1mm,label.angle=110}{v2}
			\fmfdot{v1,v2}
		\end{fmfgraph*}
		}
		\\[2mm]
		&\hspace{2mm}+\hspace{2mm}
		\raisebox{-2mm}{
		\begin{fmfgraph*}(34.2,5.4)\fmfstyle
			\fmfleft{o}\fmfright{i}
			\fmfpoly{full}{p11,p12,p13,p14}
			\fmfpoly{shaded}{p21,p22,p23,p24}
			\fmf{wiggly}{v1,v2}
			\fmffixedy{h}{p11,p12}\fmffixedy{h}{p21,p22}\fmffixedx{16}{v1,v2}
			\fmf{phantom}{i,p21}\fmf{phantom}{p24,v2}\fmf{phantom}{v2,p23}\fmf{phantom}{p22,i}
			\fmf{phantom}{v1,p11}\fmf{phantom}{p14,o}\fmf{phantom}{o,p13}\fmf{phantom}{p12,v1}
			\fmffreeze
			\fmfi{fermion}{vloc(__o){1,2} .. tension 1 .. {right}vloc(__p13)}
			\fmfi{fermion}{vloc(__p12){right} .. tension 1 .. {1,-2}vloc(__v1)}
			\fmfi{fermion}{vloc(__v1){-1,-2} .. tension 1 .. {left}vloc(__p11)}
			\fmfi{fermion}{vloc(__p14){left} .. tension 1 .. {-1,2}vloc(__o)}
			\fmfi{fermion}{vloc(__v2){1,2} .. tension 1 .. {right}vloc(__p23)}
			\fmfi{fermion}{vloc(__p22){right} .. tension 1 .. {1,-2}vloc(__i)}
			\fmfi{fermion}{vloc(__i){-1,-2} .. tension 1 .. {left}vloc(__p21)}
			\fmfi{fermion}{vloc(__p24){left} .. tension 1 .. {-1,2}vloc(__v2)}
			\fmfdot{p11,p12,p13,p14,p21,p22,p23,p24}
			\fmfv{label=${\scriptstyle 1}$,label.dist=0.45mm}{o}
			\fmfv{label=${\scriptstyle 2}$,label.dist=0.45mm}{i}
			\fmfv{label=${\scriptstyle 2}$,label.dist=1mm,label.angle=70}{v1}
			\fmfv{label=${\scriptstyle 1}$,label.dist=1mm,label.angle=110}{v2}
			\fmfdot{v1,v2}
		\end{fmfgraph*}
		}
		\\[2mm]
		&\hspace{2mm}+\hspace{2mm}
		\raisebox{-2mm}{
		\begin{fmfgraph*}(34.2,5.4)\fmfstyle
			\fmfleft{o}\fmfright{i}
			\fmfpoly{full}{p11,p12,p13,p14}
			\fmfpoly{shaded}{p21,p22,p23,p24}
			\fmf{wiggly}{v1,v2}
			\fmffixedy{h}{p11,p12}\fmffixedy{h}{p21,p22}\fmffixedx{16}{v1,v2}
			\fmf{phantom}{i,p21}\fmf{phantom}{p24,v2}\fmf{phantom}{v2,p23}\fmf{phantom}{p22,i}
			\fmf{phantom}{v1,p11}\fmf{phantom}{p14,o}\fmf{phantom}{o,p13}\fmf{phantom}{p12,v1}
			\fmffreeze
			\fmfi{fermion}{vloc(__o){1,2} .. tension 1 .. {right}vloc(__p13)}
			\fmfi{fermion}{vloc(__p12){right} .. tension 1 .. {1,-2}vloc(__v1)}
			\fmfi{fermion}{vloc(__v1){-1,-2} .. tension 1 .. {left}vloc(__p11)}
			\fmfi{fermion}{vloc(__p14){left} .. tension 1 .. {-1,2}vloc(__o)}
			\fmfi{fermion}{vloc(__v2){1,2} .. tension 1 .. {right}vloc(__p23)}
			\fmfi{fermion}{vloc(__p22){right} .. tension 1 .. {1,-2}vloc(__i)}
			\fmfi{fermion}{vloc(__i){-1,-2} .. tension 1 .. {left}vloc(__p21)}
			\fmfi{fermion}{vloc(__p24){left} .. tension 1 .. {-1,2}vloc(__v2)}
			\fmfdot{p11,p12,p13,p14,p21,p22,p23,p24}
			\fmfv{label=${\scriptstyle 1}$,label.dist=0.45mm}{o}
			\fmfv{label=${\scriptstyle 2}$,label.dist=0.45mm}{i}
			\fmfv{label=${\scriptstyle 2}$,label.dist=1mm,label.angle=70}{v1}
			\fmfv{label=${\scriptstyle 2}$,label.dist=1mm,label.angle=110}{v2}
			\fmfdot{v1,v2}
		\end{fmfgraph*}
		}
	\end{align*}
This shows that $\chi_{12}$ is nonzero even if $\Pi_{12}=0$: in that case, the second and third terms on the right-hand side survive.

The interplate polarizabilities $\Pi_{12}$ and $\Pi_{21}$ are in general nonzero and different. In the RPA, however, they vanish. The RPA replaces the polarizability by its value in the absence of Coulomb interaction:
	\begin{equation}
		\raisebox{-2mm}{
		\begin{fmfgraph*}(16,6)\fmfstyle
			\fmfleft{o}\fmfright{i}
			\fmfpoly{full,label=\scalebox{0.7}{\textcolor{white}{\textbf{\textsf{RPA}}}}}{p1,p2,p3,p4}
			\fmffixedy{h}{p1,p2}
			\fmf{phantom}{i,p1}\fmf{phantom}{p4,o}\fmf{phantom}{o,p3}\fmf{phantom}{p2,i}
			\fmffreeze
			\fmfi{fermion}{vloc(__o){1,2} .. tension 1 .. {right}vloc(__p3)}
			\fmfi{fermion}{vloc(__p2){right} .. tension 1 .. {1,-2}vloc(__i)}
			\fmfi{fermion}{vloc(__i){-1,-2} .. tension 1 .. {left}vloc(__p1)}
			\fmfi{fermion}{vloc(__p4){left} .. tension 1 .. {-1,2}vloc(__o)}
			\fmfdot{p1,p2,p3,p4}
		\end{fmfgraph*}
		}
		=
		\raisebox{-2mm}{
		\begin{fmfgraph*}(12,6)\fmfstyle
			\fmfleft{o}\fmfright{i}
			\fmffreeze
			\fmfi{fermion}{vloc(__o){1,1.5} .. tension 1 .. {1,-1.5}vloc(__i)}
			\fmfi{fermion}{vloc(__i){-1,-1.5} .. tension 1 .. {-1,1.5}vloc(__o)}
		\end{fmfgraph*}
		}
	\end{equation}
This approximation prevents any interplate polarizability, because the particle-hole pair created on one side of the diagram---that is, on a given plate---must recombine on the same plate:
	\begin{equation}
		\raisebox{-2mm}{
		\begin{fmfgraph*}(16,6)\fmfstyle
			\fmfleft{o}\fmfright{i}
			\fmfpoly{full,label=\scalebox{0.7}{\textcolor{white}{\textbf{\textsf{RPA}}}}}{p1,p2,p3,p4}
			\fmffixedy{h}{p1,p2}
			\fmf{phantom}{i,p1}\fmf{phantom}{p4,o}\fmf{phantom}{o,p3}\fmf{phantom}{p2,i}
			\fmffreeze
			\fmfi{fermion}{vloc(__o){1,2} .. tension 1 .. {right}vloc(__p3)}
			\fmfi{fermion}{vloc(__p2){right} .. tension 1 .. {1,-2}vloc(__i)}
			\fmfi{fermion}{vloc(__i){-1,-2} .. tension 1 .. {left}vloc(__p1)}
			\fmfi{fermion}{vloc(__p4){left} .. tension 1 .. {-1,2}vloc(__o)}
			\fmfdot{p1,p2,p3,p4}
			\fmfv{label=${\scriptstyle \alpha}$,label.dist=0.5mm}{o}
			\fmfv{label=${\scriptstyle \beta}$,label.dist=0.5mm}{i}
		\end{fmfgraph*}
		}
		\hspace{2mm}=\delta_{\alpha\beta}\times\hspace{2mm}
		\raisebox{-2mm}{
		\begin{fmfgraph*}(12,6)\fmfstyle
			\fmfleft{o}\fmfright{i}
			\fmffreeze
			\fmfi{fermion}{vloc(__o){1,1.5} .. tension 1 .. {1,-1.5}vloc(__i)}
			\fmfi{fermion}{vloc(__i){-1,-1.5} .. tension 1 .. {-1,1.5}vloc(__o)}
			\fmfv{label=${\scriptstyle \alpha}$,label.dist=0.5mm}{o}
			\fmfv{label=${\scriptstyle \alpha}$,label.dist=0.5mm}{i}
		\end{fmfgraph*}
		}
	\end{equation}
More formally, it follows from Eq.~(\ref{eq:chi-diag}) that $\Pi^{\mathrm{RPA}}_{\alpha\beta}\equiv\Pi^0_{\alpha\beta}=-\chi^0_{\alpha\beta}$, which vanishes for $\alpha\neq\beta$ because $n_1(\vec{q},t)$ and $n_2(-\vec{q},0)$ commute in Eq.~(\ref{eq:chinn}) if $H_{12}=0$. The RPA leads, in the long-wavelength limit, to the Thomas--Fermi approximation in which the polarizability is proportional to the Fermi-level density of states. Applying the diagrammatic rules, one indeed finds
	\begin{align}\label{eq:RPA}
		\nonumber
		\Pi_{\alpha\alpha}^{\mathrm{RPA}}(\vec{q},\varepsilon)&=
		\hspace{2mm}\raisebox{-2mm}{
		\begin{fmfgraph*}(12,6)\fmfstyle
			\fmfleft{o}\fmfright{i}
			\fmffreeze
			\fmfi{fermion}{vloc(__o){1,1.5} .. tension 1 .. {1,-1.5}vloc(__i)}
			\fmfi{fermion}{vloc(__i){-1,-1.5} .. tension 1 .. {-1,1.5}vloc(__o)}
			\fmfv{label=${\scriptstyle \alpha}$,label.dist=0.5mm}{o}
			\fmfv{label=${\scriptstyle \alpha}$,label.dist=0.5mm}{i}
		\end{fmfgraph*}
		}\hspace{2mm}
		=-\sum_{\vec{k}\sigma}\frac{f(\xi_{\alpha\vec{k}})-f(\xi_{\alpha\vec{k}+\vec{q}})}
		{\varepsilon+\xi_{\alpha\vec{k}}-\xi_{\alpha\vec{k}+\vec{q}}+i0^+}\\
		&=\nu_{\alpha}\qquad(\varepsilon=0,\,q\to0,\,T=0).
	\end{align}
In this expression, $\xi_{\alpha\vec{k}}=\varepsilon_{\alpha\vec{k}}-\mu_{\alpha}$ is the electron dispersion in plate $\alpha$ measured from the chemical potential and $f(\xi)$ is the Fermi distribution function. 

Any diagram with a continuous line connecting the left and right vertices, irrespective of how complex it is, belongs to the intraplate polarizability. Along one such line, the plate ``flavor'' is conserved like in the example below:
	\begin{equation*}
		\raisebox{-6mm}{
		\begin{fmfgraph*}(16,6)\fmfstyle
			\fmfleft{o}\fmfright{i}
			\fmffixedx{w/4}{o,v1}\fmffixedx{w/4}{v1,v2}\fmffixedx{w/4}{v2,v3}\fmffixedx{w/4}{v3,i}
			\fmffixedy{0.68h}{o,v1}\fmffixedy{0.83h}{o,v2}\fmffixedy{0.68h}{o,v3}
			\fmffixedy{0.6h}{p1,v1}\fmffixedy{0.6h}{p2,v2}\fmffixedy{0.6h}{p3,v3}
			\fmffixedx{-0.1w}{p1,v1}\fmffixedx{0.1w}{p3,v3}
			\fmf{wiggly}{p1,v1}\fmf{wiggly}{p2,v2}\fmf{wiggly}{p3,v3}
			\fmffreeze
			\fmfi{fermion,label=${\scriptstyle \alpha}$,label.dist=0.8mm}
				{vloc(__o){1,2} .. tension 1 .. {2,1}vloc(__v1)}
			\fmfi{fermion,label=${\scriptstyle \alpha}$,label.dist=1mm}
				{vloc(__v1){2,1} .. tension 1 .. {1,0}vloc(__v2)}
			\fmfi{fermion,label=${\scriptstyle \alpha}$,label.dist=1mm}
				{vloc(__v2){1,0} .. tension 1 .. {2,-1}vloc(__v3)}
			\fmfi{fermion,label=${\scriptstyle \alpha}$,label.dist=0.8mm}
				{vloc(__v3){2,-1} .. tension 1 .. {1,-2}vloc(__i)}
			\fmfdot{v1,v2,v3}
		\end{fmfgraph*}
		}
	\end{equation*}
The interplate polarizability diagrams must therefore have particle-hole pair creations and annihilations on each plate independently, and the pairs on both plates interacting via the interplate interaction. In other words, there can not be any exchange of particles between the plates (our Hamiltonian excludes tunneling between the plates). The only first-order diagram satisfying this property is
	\begin{equation*}
		\begin{fmfgraph*}(32,6)\fmfstyle
			\fmfleft{o}\fmfright{i}\fmf{wiggly}{v1,v2}
			\fmffixedx{w/3}{o,v1}\fmffixedx{w/3}{v1,v2}\fmffixedx{w/3}{v2,i}
			\fmffixedy{0}{o,v1}\fmffixedy{0}{o,v2}
			\fmffreeze
			\fmfi{fermion}{vloc(__o){1,1.5} .. tension 1 .. {1,-1.5}vloc(__v1)}
			\fmfi{fermion}{vloc(__v1){-1,-1.5} .. tension 1 .. {-1,1.5}vloc(__o)}
			\fmfi{fermion}{vloc(__v2){1,1.5} .. tension 1 .. {1,-1.5}vloc(__i)}
			\fmfi{fermion}{vloc(__i){-1,-1.5} .. tension 1 .. {-1,1.5}vloc(__v2)}
			\fmfv{label=${\scriptstyle 1}$,label.dist=0.5mm}{o}
			\fmfv{label=${\scriptstyle 2}$,label.dist=0.5mm}{i}
			\fmfv{label=${\scriptstyle 1}$,label.dist=1mm,label.angle=70}{v1}
			\fmfv{label=${\scriptstyle 2}$,label.dist=1mm,label.angle=110}{v2}
			\fmfdot{v1,v2}
		\end{fmfgraph*}
	\end{equation*}
However, this diagram is reducible and does not belong to the polarizability. Following the same idea at second order, we get two Aslamazov--Larkin-type irreducible diagrams that contribute to the interplate polarizability:
	\begin{equation}\label{eq:AL}
		\raisebox{-2mm}{
		\begin{fmfgraph*}(16,6)\fmfstyle
			\fmfleft{o}\fmfright{i}
			\fmfpoly{phantom}{p1,p2,p3,p4}
			\fmf{fermion}{p3,p4}\fmf{fermion}{p1,p2}\fmf{wiggly}{p2,p3}\fmf{wiggly}{p1,p4}
			\fmffixedy{h}{p1,p2}
			\fmf{phantom}{i,p1}\fmf{phantom}{p4,o}\fmf{phantom}{o,p3}\fmf{phantom}{p2,i}
			\fmffreeze
			\fmfi{fermion}{vloc(__o){1,2} .. tension 1 .. {right}vloc(__p3)}
			\fmfi{fermion}{vloc(__p2){right} .. tension 1 .. {1,-2}vloc(__i)}
			\fmfi{fermion}{vloc(__i){-1,-2} .. tension 1 .. {left}vloc(__p1)}
			\fmfi{fermion}{vloc(__p4){left} .. tension 1 .. {-1,2}vloc(__o)}
			\fmfdot{p1,p2,p3,p4}
			\fmfv{label=${\scriptstyle 1}$,label.dist=0.5mm}{o}
			\fmfv{label=${\scriptstyle 2}$,label.dist=0.5mm}{i}
		\end{fmfgraph*}
		}
		\qquad\text{and}\qquad
		\raisebox{-2mm}{
		\begin{fmfgraph*}(16,6)\fmfstyle
			\fmfleft{o}\fmfright{i}
			\fmfpoly{phantom}{p1,p2,p3,p4}
			\fmf{fermion}{p3,p4}\fmf{fermion}{p1,p2}\fmf{wiggly}{p2,p4}\fmf{wiggly}{p1,p3}
			\fmffixedy{h}{p1,p2}
			\fmf{phantom}{i,p1}\fmf{phantom}{p4,o}\fmf{phantom}{o,p3}\fmf{phantom}{p2,i}
			\fmffreeze
			\fmfi{fermion}{vloc(__o){1,2} .. tension 1 .. {right}vloc(__p3)}
			\fmfi{fermion}{vloc(__p2){right} .. tension 1 .. {1,-2}vloc(__i)}
			\fmfi{fermion}{vloc(__i){-1,-2} .. tension 1 .. {left}vloc(__p1)}
			\fmfi{fermion}{vloc(__p4){left} .. tension 1 .. {-1,2}vloc(__o)}
			\fmfdot{p1,p2,p3,p4}
			\fmfv{label=${\scriptstyle 1}$,label.dist=0.5mm}{o}
			\fmfv{label=${\scriptstyle 2}$,label.dist=0.5mm}{i}
		\end{fmfgraph*}
		}
	\end{equation}
Similar diagrams occur in the theory of the Coulomb drag between two parallel two-dimensional conductors \cite{[See ][ and references therein.]Carrega-2012, Narozhny-2016}. These two diagrams lead to geometric series that can be approximately evaluated in closed form. We present this evaluation below, as it shows that the corresponding polarization is proportional to the DOS in the two plates and may change sign. A more thorough investigation of $\Pi_{12}$ goes beyond the scope of the present work.

We replace in the first diagram of (\ref{eq:AL}) the bare interaction by the screened interaction using Eq.~(\ref{eq:screening-diagrams}) and include higher-order terms, obtaining the infinite series
	\begin{equation}
		\Pi_{12}^{(1)}=
		\hspace{2mm}\raisebox{-2mm}{
		\begin{fmfgraph*}(16,6)\fmfstyle
			\fmfleft{o}\fmfright{i}
			\fmfpoly{phantom}{p1,p2,p3,p4}
			\fmf{fermion}{p3,p4}\fmf{fermion}{p1,p2}\fmf{dbl_wiggly}{p2,p3}\fmf{dbl_wiggly}{p1,p4}
			\fmffixedy{h}{p1,p2}
			\fmf{phantom}{i,p1}\fmf{phantom}{p4,o}\fmf{phantom}{o,p3}\fmf{phantom}{p2,i}
			\fmffreeze
			\fmfi{fermion}{vloc(__o){1,2} .. tension 1 .. {right}vloc(__p3)}
			\fmfi{fermion}{vloc(__p2){right} .. tension 1 .. {1,-2}vloc(__i)}
			\fmfi{fermion}{vloc(__i){-1,-2} .. tension 1 .. {left}vloc(__p1)}
			\fmfi{fermion}{vloc(__p4){left} .. tension 1 .. {-1,2}vloc(__o)}
			\fmfdot{p1,p2,p3,p4}
			\fmfv{label=${\scriptstyle 1}$,label.dist=0.5mm}{o}
			\fmfv{label=${\scriptstyle 2}$,label.dist=0.5mm}{i}
		\end{fmfgraph*}
		}
		\hspace{2mm}+\hspace{2mm}
		\raisebox{-2mm}{
		\begin{fmfgraph*}(16,6)\fmfstyle
			\fmfleft{o}\fmfright{i}
			\fmfpoly{phantom}{p1,p2,p3,p4}
			\fmf{fermion}{p3,p34,p4}\fmf{fermion}{p1,p12,p2}
			\fmf{dbl_wiggly}{p2,p3}\fmf{dbl_wiggly}{p1,p4}\fmf{dbl_wiggly}{p12,p34}
			\fmffixedy{h}{p1,p2}\fmffixedy{h}{p4,p3}
			\fmffixedy{h/2}{p1,p12}\fmffixedx{0}{p1,p12}
			\fmffixedy{h/2}{p4,p34}\fmffixedx{0}{p4,p34}
			\fmf{phantom}{i,p1}\fmf{phantom}{p4,o}\fmf{phantom}{o,p3}\fmf{phantom}{p2,i}
			\fmffreeze
			\fmfi{fermion}{vloc(__o){1,2} .. tension 1 .. {right}vloc(__p3)}
			\fmfi{fermion}{vloc(__p2){right} .. tension 1 .. {1,-2}vloc(__i)}
			\fmfi{fermion}{vloc(__i){-1,-2} .. tension 1 .. {left}vloc(__p1)}
			\fmfi{fermion}{vloc(__p4){left} .. tension 1 .. {-1,2}vloc(__o)}
			\fmfdot{p1,p2,p3,p4,p12,p34}
			\fmfv{label=${\scriptstyle 1}$,label.dist=0.5mm}{o}
			\fmfv{label=${\scriptstyle 2}$,label.dist=0.5mm}{i}
		\end{fmfgraph*}
		}
		\hspace{2mm}+\ldots
	\end{equation}
The ladder sum is readily evaluated if we approximate the screened interaction by its value at $Q=0$. Here, we use the imaginary-time formalism and the ``four-vector'' notation
	\begin{equation*}
		Q\equiv(\vec{q},i\Omega_n),\quad K\equiv(\vec{k},\sigma,i\omega_n),\quad
		\sum_K\equiv\sum_{\vec{k}\sigma}\frac{1}{\beta}\sum_{i\omega_n}\hspace{0.5em},
	\end{equation*}
where $i\omega_n$ and $i\Omega_n$ are fermionic and bosonic Matsubara frequencies, $1/\beta=k_{\mathrm{B}}T$, and a unit normalization area is implied. The ladder sum is
\vspace*{1em}
	\begin{align}\label{eq:ladderph}
		\nonumber
		&\raisebox{-2mm}{
		\begin{fmfgraph*}(10,6)\fmfstyle
			\fmfleft{i1,i2}\fmfright{o1,o2}\fmffixedy{h}{p1,p2}\fmffixedy{h}{p3,p4}
			\fmffixedx{h}{p1,p3}\fmffixedx{h}{p2,p4}
			\fmf{plain}{i1,p1}\fmf{plain}{p2,i2}\fmf{plain}{p3,o1}\fmf{plain}{o2,p4}
			\fmf{fermion,label=${\scriptscriptstyle K_1}$,label.dist=0.8mm,label.side=right}{p1,p3}
			\fmf{fermion,label=${\scriptscriptstyle K_1+K'-K-Q}$,label.dist=0.6mm,label.side=right}{p4,p2}
			\fmf{dbl_wiggly}{p1,p2}\fmf{dbl_wiggly}{p3,p4}
			\fmfdot{p1,p2,p3,p4}
			\fmfv{label=${\scriptscriptstyle K}$,label.angle=180}{i1}
			\fmfv{label=${\scriptscriptstyle K'-Q}$,label.angle=180}{i2}
			\fmfv{label=${\scriptscriptstyle K+Q}$,label.angle=0}{o1}
			\fmfv{label=${\scriptscriptstyle K'}$,label.angle=0}{o2}
		\end{fmfgraph*}
		}
		\hspace{1.5em}+\hspace{1.5em}
		\raisebox{-2mm}{
		\begin{fmfgraph*}(17.5,6)\fmfstyle
			\fmfleft{i1,i2}\fmfright{o1,o2}
			\fmffixedy{h}{p1,p2}\fmffixedy{h}{p3,p4}\fmffixedy{h}{p5,p6}
			\fmffixedx{h}{p1,p3}\fmffixedx{h}{p2,p4}\fmffixedx{h}{p3,p5}\fmffixedx{h}{p4,p6}
			\fmf{plain}{i1,p1}\fmf{plain}{p2,i2}\fmf{plain}{p5,o1}\fmf{plain}{o2,p6}
			\fmf{fermion,label=${\scriptscriptstyle K_1}$,label.dist=0.8mm,label.side=right}{p1,p3}
			\fmf{fermion,label=${\scriptscriptstyle K_1+K'-K-Q}$,label.dist=0.6mm,label.side=right}{p4,p2}
			\fmf{fermion,label=${\scriptscriptstyle K_2}$,label.dist=0.8mm,label.side=right}{p3,p5}
			\fmf{fermion,label=${\scriptscriptstyle K_2+K'-K-Q}$,label.dist=2.8mm,label.side=right}{p6,p4}
			\fmf{dbl_wiggly}{p1,p2}\fmf{dbl_wiggly}{p3,p4}\fmf{dbl_wiggly}{p5,p6}
			\fmfdot{p1,p2,p3,p4,p5,p6}
			\fmfv{label=${\scriptscriptstyle K}$,label.angle=180}{i1}
			\fmfv{label=${\scriptscriptstyle K'-Q}$,label.angle=180}{i2}
			\fmfv{label=${\scriptscriptstyle K+Q}$,label.angle=0}{o1}
			\fmfv{label=${\scriptscriptstyle K'}$,label.angle=0}{o2}
		\end{fmfgraph*}
		}
		\hspace{1.5em}+ \cdots\\[1em]
		\nonumber
		&\approx\left(-W_{12}\right)^2\sum_{K_1}\mathscr{G}_{1}(K_1)\mathscr{G}_{2}(K_1+K'-K-Q)\\
		\nonumber
		&\quad+\left(-W_{12}\right)^3\Big[\sum_{K_1}\mathscr{G}_{1}(K_1)
		\mathscr{G}_{2}(K_1+K'-K-Q)\Big]^2+\ldots\\
		&=\frac{W_{12}^2\zeta^0_{12}(K'-K-Q)}{1+W_{12}\zeta^0_{12}(K'-K-Q)}\\
		\nonumber
		\zeta^0_{12}(Q)&=\sum_{K}\mathscr{G}_{1}(K)\mathscr{G}_{2}(K+Q)
		=\sum_{\vec{k}\sigma}\frac{f(\xi_{1\vec{k}})-f(\xi_{2\vec{k}+\vec{q}})}
		{i\Omega_n+\xi_{1\vec{k}}-\xi_{2\vec{k}+\vec{q}}}.
	\end{align}
Although its expression ressembles that of the noninteracting susceptibility Eq.~(\ref{eq:RPA}), the quantity $\zeta^0_{12}(Q)$ introduced here is \emph{not} a charge susceptibility. The susceptibility $\chi_{12}$ describes the propagation of a density fluctuation created on plate~2 and annihilated on plate~1: we have seen above that $\chi_{12}^0$ vanishes, because a density fluctuation cannot propagate from plate~2 to plate~1 in the absence of interplate interaction. In contrast, $\zeta_{12}$ describes the propagation of pairs created by simultaneously adding a particle to plate~2 and a hole to plate~1. These pairs propagate freely in the absence of interaction, therefore $\zeta^0_{12}$ is nonzero. The following expression results for the first series of diagrams:
	\begin{multline}\label{eq:Pi121}
		\Pi_{12}^{(1)}(Q)\approx\sum_{K}\mathscr{G}_{1}(K)\mathscr{G}_{1}(K+Q)
		\sum_{K'}\mathscr{G}_{2}(K')\mathscr{G}_{2}(K'-Q)\\
		\times\frac{W_{12}^2\zeta^0_{12}(K'-K-Q)}{1+W_{12}\zeta^0_{12}(K'-K-Q)}.
	\end{multline}
From the second diagram of Eq.~(\ref{eq:AL}), we proceed similarly to generate the cooperon series:
	\begin{equation}\label{eq:cooperon}
		\Pi_{12}^{(2)}=
		\hspace{2mm}\raisebox{-2mm}{
		\begin{fmfgraph*}(16,6)\fmfstyle
			\fmfleft{o}\fmfright{i}
			\fmfpoly{phantom}{p1,p2,p3,p4}
			\fmf{fermion}{p3,p4}\fmf{fermion}{p2,p1}\fmf{dbl_wiggly}{p2,p3}\fmf{dbl_wiggly}{p1,p4}
			\fmffixedy{h}{p1,p2}
			\fmf{phantom}{i,p1}\fmf{phantom}{p4,o}\fmf{phantom}{o,p3}\fmf{phantom}{p2,i}
			\fmffreeze
			\fmfi{fermion}{vloc(__o){1,2} .. tension 1 .. {right}vloc(__p3)}
			\fmfi{fermion}{vloc(__i){-1,2} .. tension 1 .. {left}vloc(__p2)}
			\fmfi{fermion}{vloc(__p1){right} .. tension 1 .. {1,2}vloc(__i)}
			\fmfi{fermion}{vloc(__p4){left} .. tension 1 .. {-1,2}vloc(__o)}
			\fmfdot{p1,p2,p3,p4}
			\fmfv{label=${\scriptstyle 1}$,label.dist=0.5mm}{o}
			\fmfv{label=${\scriptstyle 2}$,label.dist=0.5mm}{i}
		\end{fmfgraph*}
		}
		\hspace{2mm}+\hspace{2mm}
		\raisebox{-2mm}{
		\begin{fmfgraph*}(16,6)\fmfstyle
			\fmfleft{o}\fmfright{i}
			\fmfpoly{phantom}{p1,p2,p3,p4}
			\fmf{fermion}{p3,p34,p4}\fmf{fermion}{p2,p12,p1}
			\fmf{dbl_wiggly}{p2,p3}\fmf{dbl_wiggly}{p1,p4}\fmf{dbl_wiggly}{p12,p34}
			\fmffixedy{h}{p1,p2}\fmffixedy{h}{p4,p3}
			\fmffixedy{h/2}{p1,p12}\fmffixedx{0}{p1,p12}
			\fmffixedy{h/2}{p4,p34}\fmffixedx{0}{p4,p34}
			\fmf{phantom}{i,p1}\fmf{phantom}{p4,o}\fmf{phantom}{o,p3}\fmf{phantom}{p2,i}
			\fmffreeze
			\fmfi{fermion}{vloc(__o){1,2} .. tension 1 .. {right}vloc(__p3)}
			\fmfi{fermion}{vloc(__i){-1,2} .. tension 1 .. {left}vloc(__p2)}
			\fmfi{fermion}{vloc(__p1){right} .. tension 1 .. {1,2}vloc(__i)}
			\fmfi{fermion}{vloc(__p4){left} .. tension 1 .. {-1,2}vloc(__o)}
			\fmfdot{p1,p2,p3,p4,p12,p34}
			\fmfv{label=${\scriptstyle 1}$,label.dist=0.5mm}{o}
			\fmfv{label=${\scriptstyle 2}$,label.dist=0.5mm}{i}
		\end{fmfgraph*}
		}
		\hspace{2mm}+\ldots
	\end{equation}
Like the particle-hole ladder Eq.~(\ref{eq:ladderph}), the particle-particle ladder in Eq.~(\ref{eq:cooperon}) is readily evaluated formally if the interaction is replaced by its short-range and static limit:\vspace*{1em}
	\begin{align}\label{eq:ladderpp}
		&\raisebox{-2mm}{
		\begin{fmfgraph*}(10,6)\fmfstyle
			\fmfleft{i1,i2}\fmfright{o1,o2}\fmffixedy{h}{p1,p2}\fmffixedy{h}{p3,p4}
			\fmffixedx{h}{p1,p3}\fmffixedx{h}{p2,p4}
			\fmf{plain}{i1,p1}\fmf{plain}{p2,i2}\fmf{plain}{p3,o1}\fmf{plain}{o2,p4}
			\fmf{fermion,label=${\scriptscriptstyle K_1}$,label.dist=0.8mm,label.side=right}{p1,p3}
			\fmf{fermion,label=${\scriptscriptstyle K+K'-K_1}$,label.dist=0.6mm,label.side=left}{p2,p4}
			\fmf{dbl_wiggly}{p1,p2}\fmf{dbl_wiggly}{p3,p4}
			\fmfdot{p1,p2,p3,p4}
			\fmfv{label=${\scriptscriptstyle K}$,label.angle=180}{i1}
			\fmfv{label=${\scriptscriptstyle K'}$,label.angle=180}{i2}
			\fmfv{label=${\scriptscriptstyle K+Q}$,label.angle=0}{o1}
			\fmfv{label=${\scriptscriptstyle K'-Q}$,label.angle=0}{o2}
		\end{fmfgraph*}
		}
		\hspace{1.5em}+\hspace{1.5em}
		\raisebox{-2mm}{
		\begin{fmfgraph*}(17.5,6)\fmfstyle
			\fmfleft{i1,i2}\fmfright{o1,o2}
			\fmffixedy{h}{p1,p2}\fmffixedy{h}{p3,p4}\fmffixedy{h}{p5,p6}
			\fmffixedx{h}{p1,p3}\fmffixedx{h}{p2,p4}\fmffixedx{h}{p3,p5}\fmffixedx{h}{p4,p6}
			\fmf{plain}{i1,p1}\fmf{plain}{p2,i2}\fmf{plain}{p5,o1}\fmf{plain}{o2,p6}
			\fmf{fermion,label=${\scriptscriptstyle K_1}$,label.dist=0.8mm,label.side=right}{p1,p3}
			\fmf{fermion,label=${\scriptscriptstyle K+K'-K_1}$,label.dist=0.6mm,label.side=left}{p2,p4}
			\fmf{fermion,label=${\scriptscriptstyle K_2}$,label.dist=0.8mm,label.side=right}{p3,p5}
			\fmf{fermion,label=${\scriptscriptstyle K+K'-K_2}$,label.dist=2.8mm,label.side=left}{p4,p6}
			\fmf{dbl_wiggly}{p1,p2}\fmf{dbl_wiggly}{p3,p4}\fmf{dbl_wiggly}{p5,p6}
			\fmfdot{p1,p2,p3,p4,p5,p6}
			\fmfv{label=${\scriptscriptstyle K}$,label.angle=180}{i1}
			\fmfv{label=${\scriptscriptstyle K'}$,label.angle=180}{i2}
			\fmfv{label=${\scriptscriptstyle K+Q}$,label.angle=0}{o1}
			\fmfv{label=${\scriptscriptstyle K'-Q}$,label.angle=0}{o2}
		\end{fmfgraph*}
		}
		\hspace{1.5em}+ \cdots\\[1em]
		\nonumber
		&\approx\frac{W_{12}^2\bar{\zeta}^0_{12}(K+K')}{1+W_{12}\bar{\zeta}^0_{12}(K+K')}\\
		\nonumber
		\bar{\zeta}^0_{12}(Q)&=\sum_{K}\mathscr{G}_1(K)\mathscr{G}_2(Q-K)
		=\sum_{\vec{k}\sigma}\frac{f(\xi_{1\vec{k}})-f(-\xi_{2\vec{q}-\vec{k}})}
		{i\Omega_n-\xi_{1\vec{k}}-\xi_{2\vec{q}-\vec{k}}}.
	\end{align}
The function $\bar{\zeta}^0_{12}$ describe the propagation of pairs created by simultaneously adding a particle to plates 1 and 2. This quantity is ultraviolet divergent if the dispersions $\xi_{\alpha\vec{k}}$ are unbounded. The divergence arises from ignoring a natural cutoff in $W_{12}(Q)$ at large $Q$. It may be removed by restoring a cutoff in the momentum integrals. Alternatively, a standard way of curing the divergence \cite{Huang-1957, Galitskii-1958, Bloom-1975, Nozieres-1985, Randeria-1989, Perali-2004} is to express the full ladder [including the term of first order in $W_{12}$ that is not present in Eq.~(\ref{eq:ladderpp})] in terms of the two-particle t-matrix, which contains information about the interplate bound state. Both approaches yield equivalent results and we first adopt the t-matrix solution here. We will revert to the cutoff later when discussing the sign of the cross-polarizability. For a contact interaction $W_{12}$, the t-matrix is given by
	\begin{equation}\label{eq:t-matrix}
		T^{-1}_{12}(\vec{q},E)=W^{-1}_{12}-\sum_{\vec{k}\sigma}
		\frac{1}{E-\frac{\hbar^2q^2}{2M}-\frac{\hbar^2k^2}{2\mu}+i0^+},
	\end{equation}
where $M=m_1+m_2$ and $\mu=m_1m_2/(m_1+m_2)$ are the total and reduced masses, respectively. The full particle-particle ladder---often called the pseudopotential and denoted $\Gamma$---obeys a Bethe-Salpeter equation analogous to Eq.~(\ref{eq:screening-diagrams}): $-\Gamma=-W+(-W)\bar{\zeta}^0(-\Gamma)$. The partial series in Eq.~(\ref{eq:ladderpp}) corresponds to $W-\Gamma$. Using the t-matrix in order to eliminate $W$ from the Bethe-Salpeter equation, one arrives at the following exact representation of the pseudopotential:
	\begin{multline*}
		\Gamma^{-1}_{12}(Q)=T^{-1}_{12}(\vec{q},E)\\
		-\sum_{\vec{k}\sigma}\left[\frac{1-f(\xi_{1\vec{k}})-f(\xi_{2\vec{q}-\vec{k}})}
		{i\Omega_n-\xi_{1\vec{k}}-\xi_{2\vec{q}-\vec{k}}}
		-\frac{1}{E-\frac{\hbar^2q^2}{2M}-\frac{\hbar^2k^2}{2\mu}+i0^+}\right].
	\end{multline*}
The momentum sum is now convergent, because $\xi_{1\vec{k}}+\xi_{2\vec{q}-\vec{k}}$ approaches $\hbar^2k^2/(2\mu)$ at large $k$. There is freedom in choosing $\vec{q}$ and $E$ in the t-matrix. The t-matrix diverges at the bound state energy, which occurs below the minimum of the two-particle continuum given by $\hbar^2q^2/(2M)$. We can write the bound-state energy as $E_b(\vec{q})=\hbar^2q^2/(2M)-|E_b|$, such that $T^{-1}_{12}\left(\vec{q},E_b(\vec{q})\right)=0$. With the choice $E=E_b(\vec{q})$, the pseudopotential reads
	\begin{equation}
		\Gamma^{-1}_{12}(Q)=
		\sum_{\vec{k}\sigma}\left[\frac{f(\xi_{1\vec{k}})+f(\xi_{2\vec{q}-\vec{k}})-1}
		{i\Omega_n-\xi_{1\vec{k}}-\xi_{2\vec{q}-\vec{k}}}
		-\frac{1}{|E_b|+\frac{\hbar^2k^2}{2\mu}}\right].
	\end{equation}
The series of diagrams Eq.~(\ref{eq:cooperon}) yields an expression like Eq.~(\ref{eq:Pi121}), where the fraction at the second line is replaced by $W_{12}-\Gamma_{12}(K+K')$. Collecting both series, we arrive at the expression
	\begin{multline}\label{eq:twoseries}
		\Pi_{12}^{(1+2)}(Q)\approx\sum_{K}\mathscr{G}_{1}(K)\mathscr{G}_{1}(K+Q)
		\sum_{K'}\mathscr{G}_{2}(K')\mathscr{G}_{2}(K'-Q)\\
		\times\left[\frac{W_{12}^2\zeta^0_{12}(K'-K-Q)}{1+W_{12}\zeta^0_{12}(K'-K-Q)}
		+W_{12}-\Gamma_{12}(K+K')\right].
	\end{multline}
Due to the $Q$ dependencies of $\zeta^0_{12}(Q)$ and $\Gamma_{12}(Q)$, progress is difficult without making further approximations. In line with the neglect of the $Q$ dependence of $W_{12}$, we may ignore the $Q$ dependencies of the effective interactions represented by $\zeta^0_{12}$ and $\Gamma_{12}$. The first two factors in the right-hand side of Eq.~(\ref{eq:twoseries}) become $-\Pi_{11}^{\mathrm{RPA}}(Q)$ and $-\Pi_{22}^{\mathrm{RPA}}(-Q)$, respectively, such that for $Q=0$ these two factors reduce to $\nu_1\nu_2$ [see Eq.~(\ref{eq:RPA})]. We are left with the following expression for the long-wavelength cross-polarizability:
	\begin{equation}\label{eq:Pi1212}
		\Pi_{12}^{(1+2)}(0)\approx\nu_1\nu_2\left[\frac{W_{12}^2\zeta^0_{12}(0)}{1+W_{12}\zeta^0_{12}(0)}
		+W_{12}-\Gamma_{12}(0)\right].
	\end{equation}
Looking back at Eq.~(\ref{eq:AL}), we recognize that the factors $\nu_1$ and $\nu_2$ arise from the loops on the left and right, respectively, while the terms in the square brackets give the effective interplate interactions at  long wavelength in the particle-hole and particle-particle channels. Both interactions are of order $W_{12}^2$, such that their signs are not set by the sign of the screened interaction $W_{12}$. In the particle-hole channel, the sign of the effective interaction is the sign of $\zeta^0_{12}(0)$. In the particle-particle channel, $\Gamma_{12}(0)$ approaches $W_{12}$ for $W_{12}\to0$ such that $W_{12}-\Gamma_{12}(0)\sim W_{12}^2$ and the sign of the effective interaction is fixed by the second derivative $d^2\Gamma_{12}/dW_{12}^2$.

We now evaluate $\zeta^0_{12}(0)$ and $\Gamma_{12}(0)$ at zero temperature for two parabolic bands $\xi_{\alpha\vec{k}}=\hbar^2k^2/(2m_{\alpha})-\mu_{\alpha}$. Since $\xi_{2\vec{k}}$ can be expressed in terms of $\xi_{1\vec{k}}$ and vice-versa, the momentum sums can be rewritten as integrals involving the DOS of plates $1$ and $2$:
	\begin{align*}
		\zeta^0_{12}(0)&=\int_{-\infty}^{\infty}d\xi\,\frac{\nu_1(\xi)f(\xi)}
		{\xi\left(1-\frac{m_1}{m_2}\right)-\frac{m_1}{m_2}\mu_1+\mu_2+i0^+}\\
		&\quad-\int_{-\infty}^{\infty}d\xi\,\frac{\nu_2(\xi)f(\xi)}
		{\xi\left(\frac{m_2}{m_1}-1\right)+\frac{m_2}{m_1}\mu_2-\mu_1+i0^+}.
	\end{align*}
The plate DOS are constant as soon as $\xi>-\mu_{1,2}$:
	\begin{align*}
		m_2\nu_1(\xi)&=\frac{m_1m_2}{\pi\hbar^2}\theta(\xi+\mu_1)\\
		m_1\nu_2(\xi)&=\frac{m_1m_2}{\pi\hbar^2}\theta(\xi+\mu_2),
	\end{align*}
such that
	\begin{align}\label{eq:zeta12}
		\nonumber
		\zeta^0_{12}(0)&=\frac{m_1m_2}{\pi\hbar^2}\left[\int_{-\mu_1}^0\frac{d\xi}
		{\xi\left(m_2-m_1\right)-m_1\mu_1+m_2\mu_2+i0^+}\right.\\
		\nonumber
		&\quad\left.-\int_{-\mu_2}^0\frac{d\xi}
		{\xi\left(m_2-m_1\right)+m_2\mu_2-m_1\mu_1+i0^+}\right]\\
		&=-\frac{m_1m_2}{\pi\hbar^2(m_1-m_2)}\ln\left(\frac{m_1}{m_2}\right).
	\end{align}
Clearly $\zeta^0_{12}(0)<0$. In particular, $\zeta^0_{12}(0)$ approaches the intraplate susceptibility $\chi^0_{11}=-m_1/(\pi\hbar^2)$ in the limit where $m_2$ approaches $m_1$. Proceeding similarly and using the relation $\mu_{\alpha}=\pi\hbar^2n_{\alpha}/m_{\alpha}$ with $n_{\alpha}$ the density in plate $\alpha$, we find
	\begin{equation}\label{eq:Gammainv}
		\Gamma^{-1}_{12}(0)=\frac{\mu}{\pi\hbar^2}
		\ln\left[\frac{|E_b|}{\pi\hbar^2}\frac{m_2n_1+m_1n_2}{(n_1-n_2)^2}\right]
	\end{equation}
for the pseudopotential, with $\mu$ the reduced mass. Note that $n_1$ and $n_2$ are the equilibrium densities in the plates. It appears that the effective interaction in the particle-particle channel approaches zero logarithmically in the particular case of two plates with the same density. In order to determine the sign of the effective interaction, it is necessary to substitute in Eq.~(\ref{eq:Gammainv}) $|E_b|$ by its expression in terms of the bare interaction $W_{12}$. Since this removes $|E_b|$ from the expressions, it would restore the ultraviolet singularity unless a high-energy cutoff is simultaneously introduced. To this end, we solve Eq.~(\ref{eq:t-matrix}) for the bound state, cutting the divergence at energy $E_c$:
	\begin{equation*}
		\frac{1}{|W_{12}|}=\frac{\mu}{\pi\hbar^2}\int_0^{E_c}
		\frac{d\varepsilon}{|E_b|+\varepsilon-i0^+}=\frac{\mu}{\pi\hbar^2}
		\ln\left(1+\frac{E_c}{|E_b|}\right),
	\end{equation*}
which gives the solution
	\begin{equation}\label{eq:Eb}
		|E_b|=\frac{E_c}{\exp\left(\frac{\pi\hbar^2}{\mu|W_{12}|}\right)-1}
		\approx E_c \exp\left(-\frac{\pi\hbar^2}{\mu|W_{12}|}\right),
	\end{equation}
the last expression being accurate for small $W_{12}$. Inserting Eq.~(\ref{eq:Eb}) in Eq.~(\ref{eq:Gammainv}) yields
	\begin{equation}\label{eq:Gamma}
		W_{12}-\Gamma_{12}(0)=\frac{W_{12}^2}{W_{12}+\frac{\pi\hbar^2}{\mu}/
		\ln\left[\frac{E_c}{\pi\hbar^2}\frac{m_2n_1+m_1n_2}{(n_1-n_2)^2}\right]}.
	\end{equation}
As the bare interplate potential is cut at short wavelength by a factor $e^{-qd}$, the screened potential must behave similarly and we can use $\hbar^2/(\mu d^2)$ as the cutoff. Introducing this, Eq.~(\ref{eq:zeta12}), and Eq.~(\ref{eq:Gamma}) in Eq.~(\ref{eq:Pi1212}) and keeping the lowest order in $W_{12}$ leads to our final result reported in the main text, Eq.~(\ref{eq:Pi12}).

\end{fmffile}

\section{Quantum capacitance and local-field factors}
\label{app:STLS}

The STLS theory \cite{Singwi-1968} includes quantum effects in the susceptibility by means of a local-field factor $G(q)$ instead of the polarizability. The role of $G(q)$ is to take into account correlations in the screening process. The primary response to the external potential (see Appendix~\ref{app:chiclass}) is taken as $\chi_0U$, like in the RPA. The response to the secondary potential $V\chi_0U$, however, is not $\chi_0V\chi_0U$ like in RPA, but $\chi_0(1-G)V\chi_0U$. The correction $-G$ represents the effect of the exchange-correlation hole, which changes the screening at short distances. It can be shown that $G(0)=0$, because the long-distance physics is not affected by the correlations, and that $\lim_{q\to\infty}[1-G(q)]=g(0)$, where $g(r)$ is the pair-distribution function \cite{Mahan}. The STLS theory provides a self-consistent definition of $G(q)$ that can be solved numerically. Once generalized to the geometry of the capacitor \cite{Zheng-1994}, it provides another way of expressing the quantum capacitance. The susceptibility matrix of the capacitor now takes the form
	\begin{equation}\label{eq:chi-G}
		\chi^{-1}=\begin{pmatrix}\chi_{01}^{-1}-V_{11}(1-G_{11}) & V_{12}(G_{12}-1)\\
		V_{21}(G_{21}-1) & \chi_{02}^{-1}-V_{22}(1-G_{22}) \end{pmatrix},
	\end{equation}
where $\chi_{0\alpha}(q)$ are the noninteracting susceptibilities and $G_{\alpha\beta}(q)$ are the local-field factors. By comparing Eq.~(\ref{eq:chi-G}) with Eq.~(\ref{eq:chi-diag}), one can express $\Pi_{\alpha\beta}$ in terms of $G_{\alpha\beta}$, $V_{\alpha\beta}$, and $\chi_{0\alpha}$. After substitution into Eq.~(\ref{eq:C-Pi}), the result for the quantum capacitance is
	\begin{equation*}\label{eq:C-G0}
		\frac{e^2}{\CQ}=\lim_{q\to0}\left[-\frac{1}{\chi_{01}}-\frac{1}{\chi_{02}}-(G_{11}+G_{22})V_{11}
		-(G_{12}+G_{21})V_{12}\right].
	\end{equation*}
Because it rests on the noninteracting susceptibilities, this representation singles out the kinetic capacitances $C_{\mathrm{kin},\alpha}/e^2=\nu_{\alpha}=-\chi_{0\alpha}$. This leads to the expression
	\begin{align}\label{eq:C-G}
		\frac{1}{\CQ}&=\frac{1}{\Ck}+\frac{1}{C_{\mathrm{kin},2}}-\frac{1}{\Delta C}\\
		\nonumber
		\frac{1}{\Delta C}&=
		\frac{1}{2\epsilon}\lim_{q\to0}\frac{G_{11}(q)+G_{22}(q)-e^{-qd}[G_{12}(q)+G_{21}(q)]}{q}.
	\end{align}
The RPA corresponds to $G_{\alpha\beta}=0$ and only gives the kinetic capacitances, as found previously. The local-field factors calculated for electron-hole bilayers \cite{Liu-1996, Dong-1998, Moudgil-2002} vanish linearly for $q\to0$ with a positive slope for the diagonal elements and a negative slope for the off-diagonal ones. Hence $\Delta C>0$ and $\CQ>\Ck/2$ for two identical plates, consistently with Fig.~\ref{fig:identical-plates}.

\section{Comparison of Eqs.~(\ref{eq:model1}) and (\ref{eq:Pid})}
\label{app:comparison}

We compare here the model of quantum capacitance derived in Sec.~\ref{sec:symmetric} for a symmetric capacitor with the diagrammatic result. To this end, we express the quantum capacitance given by Eq.~(\ref{eq:model1}) in terms of the same dimensionless variables as used in Eq.~(\ref{eq:Pid}), $x=d/a_1$ and $y=(8\pi n_1a_1^2)^{1/2}$, and we normalize it by the kinetic capacitance of plate~1, which yields
	\begin{equation}\label{eq:CQ1}
		\frac{\CQ^{(\mathrm{I})}}{\Ck}
		=\frac{y/2}{y-1}+\frac{1}{4x}\left(\sqrt{1+\frac{2xy}{y-1}}-1\right).
	\end{equation}
The condition of validity $d>\ell_1/2$ translates into $x>\frac{1}{2}(1-1/y)$. We would like to compare this expression with the diagrammatic result, Eq.~(\ref{eq:Pid}), specialized for identical plates, i.e., with $u=1$ and $v=1$. Because Eq.~(\ref{eq:model1}) requires $\Pi_{12}\ll\Pi_{11}$, we use the non self-consistent model by setting $\Pi_{12}$ to zero in Eq.~(\ref{eq:Pidc}). Equation~(\ref{eq:Pidb}) is regular at $u=1$ but has a singularity at $v=1$. As $\Pi_{12}$ depends only logarithmically on $v$ close to the singularity, we set $v=1.00001$ to represent plates with equal densities. The resulting quantum capacitance is
	\begin{multline}\label{eq:CQd}
		\frac{\CQ^{(\mathrm{II})}}{\Ck}
		=\frac{y/2}{y-1}+\frac{(y-1)^4}{16y^2(y-1+xy)^2}\\
		\times\left\{-2+\ln\left[\frac{8}{(1-v)^2(xy)^2}\right]\right\}.
	\end{multline}

\begin{figure}[b]
\includegraphics[width=\columnwidth]{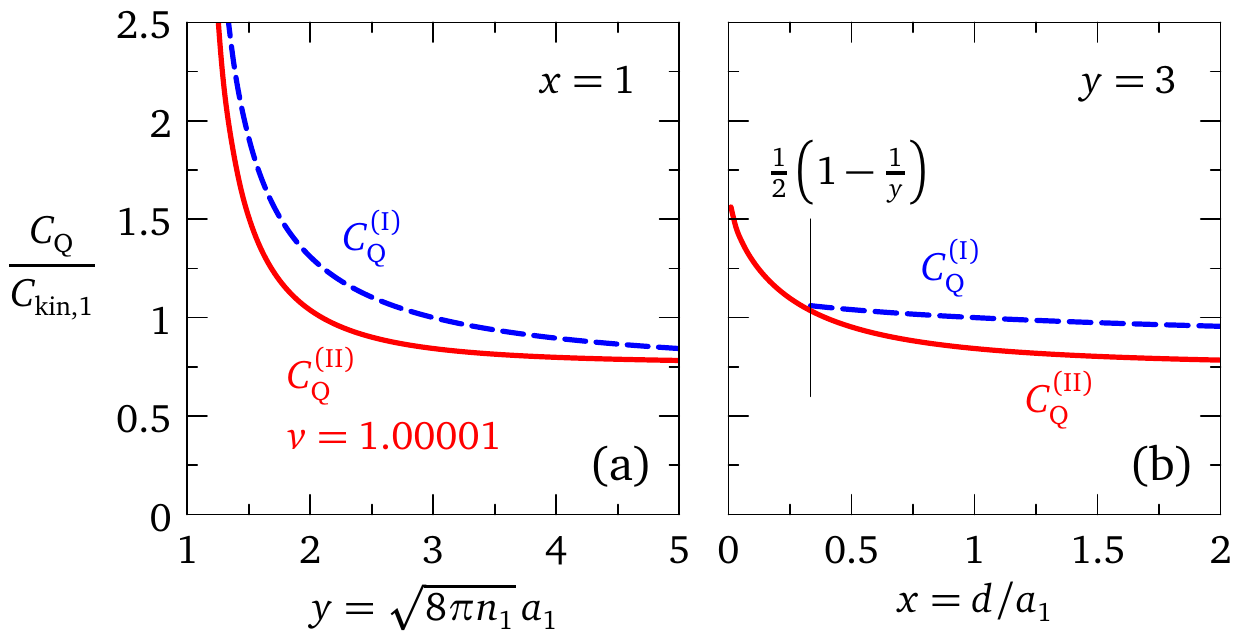}
\caption{\label{fig:comparison}
Comparison of the two quantum capacitance models for a symmetric capacitor. The capacitance is plotted (a) vs density for dimensionless thickness $x=1$ and (b) vs thickness for dimensionless density $y=3$. The dashed lines show the model based on screening lengths, Eq.~(\ref{eq:CQ1}), which is only valid in the range $x>\frac{1}{2}(1-1/y)$. The solid lines show the diagrammatic result, Eq.~(\ref{eq:CQd}).
}
\end{figure}

Figure~\ref{fig:comparison} compares the two models. They both yield similar orders of magnitude and similar trends for the quantum capacitance versus density and thickness. The cross-polarizability is positive in this configuration of capacitor and the quantum capacitance is increased relative to the RPA value, diverging a low density similarly in both models [Fig.~\ref{fig:comparison}(a)]. For very small thicknesses, the capacitance $\CQ^{(\mathrm{II})}$ diverges logarithmically due to the singular contribution of the bound state. The model $\CQ^{(\mathrm{I})}$ is not applicable in this regime, where $x<\frac{1}{2}(1-1/y)$ [Fig.~\ref{fig:comparison}(b)].

\end{document}